\documentclass[sigplan,10pt]{acmart}
\acmISBN{} 
\acmDOI{} 
\startPage{1}

\setcopyright{none}

\usepackage{booktabs}   
\usepackage{subcaption} 
\usepackage{tikz}
\usepackage{amsmath}
\usepackage{xspace}
\usepackage{textcomp}
\usepackage{enumitem}
\usepackage[noend]{algpseudocode}
\usepackage{underscore}
\usepackage{algorithm}
\usepackage{multirow}
\usepackage{multicol}
\usepackage{algorithmicx}
\usepackage{makecell}
\usepackage{colortbl}

\newcommand{\SYS}{DFOGraph\xspace}
\newcommand{\twi}{\emph{twitter-2010}\xspace}
\newcommand{\uk}{\emph{uk-2014}\xspace}
\newcommand{\rmat}{\emph{RMAT-32}\xspace}
\newcommand{\kron}{\emph{KRON-38}\xspace}
\newcommand{\processVerticesNoSpace}{\texttt{ProcessVertices}}
\newcommand{\processVertices}{\processVerticesNoSpace\xspace}
\newcommand{\processEdgesNoSpace}{\texttt{ProcessEdges}}
\newcommand{\processEdges}{\processEdgesNoSpace\xspace}
\newcommand{\process}{\texttt{Process}\xspace}

\newcommand{\darray}{\texttt{VertexArray}\xspace}
\newcommand{\darrays}{\texttt{VertexArray}s\xspace}
\newcommand{\getdarraynospace}{\texttt{GetVertexArray}}
\newcommand{\getdarray}{\getdarraynospace\xspace}

\begin{document}

\title[\SYS]{\huge \SYS: An I/O- and Communication-Efficient System for Distributed Fully-out-of-Core Graph Processing}         


\author{Jiping Yu, Wei Qin, Xiaowei Zhu, Zhenbo Sun, Jianqiang Huang, Xiaohan Li, and Wenguang Chen}
\affiliation{
  \institution{Tsinghua University}            
}        

\begin{abstract}
With the magnitude of graph-structured data continually increasing, graph processing systems that can scale-out and scale-up are needed to handle extreme-scale datasets. While existing distributed out-of-core solutions have made it possible, they suffer from limited performance due to excessive I/O and communication costs.

We present \SYS, a distributed fully-out-of-core graph processing system that applies and assembles multiple techniques to enable I/O- and communication-efficient processing. \SYS builds upon two-level column-oriented partition with adaptive compressed representations to allow fine-grained selective computation and communication, and it only issues necessary disk and network requests.
Our evaluation shows \SYS achieves performance comparable to GridGraph and FlashGraph (>2.52$\times$ and 1.06$\times$) on a single machine and outperforms Chaos and HybridGraph significantly (>12.94$\times$ and >10.82$\times$) when scaling out.
\end{abstract}



\maketitle

\begin{table*}[tb!]
  \centering
    \begin{tabular}{c|c|c|c|c|c|c}
      \toprule[1.5pt]
      Feature & \SYS & Chaos & HybridGraph & TurboGraph++ & GraphD & Gemini \\
      \midrule[1pt]
      
      \makecell{Processing\\model} &
      \makecell{Vertex-centric\\push\\signal-slot} &
      \makecell{Edge-centric\\GAS} &
      \makecell{Vertex-centric\\push \& pull\\ Pregel-like} &
      \makecell{Neighborhood-\\centric\\GAS \& NWSM} &
      \makecell{Vertex-centric\\push\\Pregel-like} &
      \makecell{Vertex-centric\\push \& pull\\signal-slot} \\
      
      \hline
      
      Out-of-core & \cellcolor{green!10} Fully-OOC &  \cellcolor{green!10}Fully-OOC & Semi-OOC & Semi-OOC & Semi-OOC & In-memory \\
      
      \hline
      \makecell{Bandwidth\\assumption} &
       \cellcolor{green!10}\makecell{Network $\ge$ \\ disk per node} &
      \makecell{Network $\ge$ \\ disk aggregated} &
       \cellcolor{green!10}\makecell{Uses bandwidth\\as tuning\\parameters} &
      \makecell{Network is not\\the bottleneck} &
       \cellcolor{green!10}\makecell{Commodity mag-\\netic disks and\\Gigabit networks} &
      \makecell{(N/A for\\in-memory\\system)} \\
      \bottomrule[1.5pt]
    \end{tabular}
  \caption{Comparison among distributed graph processing systems. Background colors indicate more advanced features.}
  \label{tbl:featcomp}
\end{table*}

\section{Introduction}

Internet-scale graphs, such as web crawls and social networks, can have hundreds of billions of vertices ($|V|$) and trillions of edges ($|E|$). Analysis of these massive graphs can be challenging, and recent works include scale-up and scale-out solutions. Scale-up ones, such as GraphChi \cite{GraphChi}, GridGraph \cite{GridGraph}, FlashGraph \cite{FlashGraph}, and many others \cite{TurboGraph,XStream,NXGraph,Graphene,Mosaic,ai2018clip}, process the graph on a single machine using a disk or array of disks.
Scale-out solutions include distributed in-memory systems, such as Pregel \cite{Pregel}, PowerGraph \cite{PowerGraph}, GraphX \cite{GraphX}, Gemini \cite{Gemini}, and many others \cite{Giraph,Hama,GraphLab,PowerSwitch,PowerLyra,Cyclops,GraM}, as well as distributed out-of-core ones, such as Chaos \cite{Chaos}, HybridGraph \cite{HybridGraph}, TurboGraph++ \cite{TurboGraph++}, and others \cite{graphd,pregelix}.

Although distributed in-memory systems have excellent performance, they require a large number of compute nodes and incur high costs. A real-world social graph \cite{backstrom2012four} consists of 721 million vertices and 137 billion edges, resulting in \char`\~1 TB raw data (assuming edges stored as pairs of 32-bit integers). A web crawl from a search engine \cite{ShenTu} contains 272 billion vertices and 12 trillion edges. The raw data exceeds 136 TB, and in-memory processing currently needs a supercomputer \cite{ShenTu}. Thus, in-memory systems suffer from high costs when processing these large graphs.

In contrast, distributed out-of-core systems are more cost-efficient. Chaos, the most well-known system in this category, can process a trillion-edge graph using 32 servers, each with 32 GB RAM. Achieving high performance is challenging, mostly caused by excessive I/O and communication traffics. Chaos needs >3 hours for each PageRank iteration on a trillion-edge graph ($|V|=2^{36},|E|=2^{40}$). On the other hand, an in-memory system GraM can finish each iteration in 140 seconds on a similar-scaled graph ($|V|=2^{33},|E|=2^{40}$).


\subsection{Needs for Fully-Out-of-Core Processing}

We would further distinguish semi- and fully-out-of-core graph processing. Semi-out-of-core systems assume vertex data can fit in memory, while fully-out-of-core ones do not. Some systems may operate fully-out-of-core but with quite limited performance. For example, GridGraph maintains vertex data using memory-mapped arrays, thus experiences excessive page swaps with insufficient memory. Experiments of TurboGraph++ use at most $|V|=2^{34}$, and the aggregate memory is 800 GB, which is far from fully-out-of-core.

We argue that it is essential to optimize for fully-out-of-core scenarios. Firstly, it is inappropriate to assume $|V| \ll |E|$. For example, in Social Network Benchmark of Linked Data Benchmark Council \cite{boncz2013ldbc}, with scale factors from 1 to 1000, the average degree is only about 6.
Semi-out-of-core systems have narrow applications on such graphs since they cannot efficiently process more massive graphs due to vertices' excessive random access on disks, and in-memory systems can better process smaller graphs.
Also, machine-learning related graph algorithms, such as node2vec \cite{grover2016node2vec}, require the data on each vertex to be vectors of, rather than one or two numeric values. Generally, a graph has $|V|<|E|$, but vertex data may be comparable to or even more extensive than edge data. 

\subsection{Challenges of Scaling Out}


The capacity and bandwidth of a single machine always limit single-node systems' capabilities to handle extreme-scale datasets.
While distributed systems can handle more massive graphs using multiple machines, they rely on fast networks to achieve scalability.
Unfortunately, some existing systems require rigorous conditions to scale out the performance effectively.
Chaos requires the network bandwidth to outstrip the \emph{aggregate} disk bandwidth of the whole cluster to scale its performance, hardly fulfilled for typical cluster environments, especially for those with fast storages\footnote{As a concrete example, each AWS EC2 i3en.3xlarge instance is equipped with an NVMe SSD of 2 GB/s (or 16 Gbps) sequential throughput and 25 Gbps network. The requirement is not satisfied even with two instances.}.
TurboGraph++ also assumes that communication is no longer a bottleneck in modern clusters with high-speed networks, which is questionable in practical configurations.


Existing distributed out-of-core systems usually focus on optimizing the I/O efficiency but pay less attention to communication costs.
While HybridGraph proposes a block-centric pull-mode propagation method to reduce the I/O and communication overhead, the actual effect depends on the memory budget available for message combining. For massive graphs far beyond the memory capacity, the reduction would be much less effective as the memory for combining is limited.

As a result, if we effectively reduce the communication cost and design a scalable system, we can relax the assumption on the network (only requires network bandwidth comparable to the disk throughput) to make distributed processing practical on more types of hardware configurations.

\subsection{Main Contributions of Our Work}

In this paper, we propose \SYS to fill the performance gap between conventional distributed in-memory and out-of-core graph processing systems.
\SYS targets cluster environments equipped with high-speed NVMe SSDs and networks, to achieve scalability on both capacity and performance and improve the overall efficiency of fully-out-of-core graph processing.
The techniques applied in \SYS focus on optimizing the I/O and communication efficiency, trying to avoid unnecessary disk and network operations, and adaptively choosing among strategies to fully utilize CPU, disk, or network whichever is the bottleneck. The main contributions of \SYS are summarized as follows:


\begin{itemize}[leftmargin=*,itemsep=0pt]
    \item The critical choice of \SYS is the combination of vertex-centric \cite{Thinking} push abstraction and two-level (inter-node and intra-node) column-oriented partitioning. Pushing makes various optimizations possible, while partitioning narrows the span of random access and makes the pushing practical in distributed fully-out-of-core scenarios without touching excessive on-disk pages.
    \item Push and two-level column-oriented partitions enable effective I/O and communication optimizations. CSR (Compressed Sparse Row) or DCSR (Doubly-Compressed Sparse Row) are adaptively chosen for edge representation of each partition, reducing the space consumption and I/O cost. Messages are efficiently filtered, and only needed ones are sent on the wire, to reduce network traffic. As a result, \SYS only requires the network bandwidth is comparable to per-node disk throughput to scale out.
    \item Moreover, \SYS develops multiple adaptive strategies of communication. Operations related to disk and network are carefully decomposed and pipelined, which mostly hides the extra latency of optimizations and helps better utilize and trade-off among CPU, network, and I/O.
    \item We compare \SYS with five state-of-the-art single-machine and distributed graph processing systems. Results show that \SYS achieves performance comparable to single-machine out-of-core systems (>2.52$\times$ over GridGraph, 1.06$\times$ over FlashGraph). 
    Though slower than Gemini (4.76$\times$) when data fits into memory, \SYS can process much larger graphs with the help of external memory.
    \SYS scales well (6.56$\times$ with eight nodes) and significantly outperforms existing distributed out-of-core systems (>12.94$\times$ over Chaos, >10.82$\times$ over HybridGraph).
\end{itemize}
Table \ref{tbl:featcomp} shows the main differences between \SYS and related distributed systems. As a fully-out-of-core one, \SYS relaxed the bandwidth assumption to scale out, because of adequate optimizations enabled by push mode and column-oriented partitions. Also, it is non-trivial to convert semi-out-of-core systems to efficient fully-out-of-core ones.

The remaining part of this paper is organized as follows. Section \ref{sec:abstraction} introduces the choice of the push abstraction and the column-oriented partitions. Section \ref{sec:implementation} introduces the implementation framework. Section \ref{sec:coreidea} discusses the I/O and communication optimizations. Section \ref{sec:evaluation} compares \SYS with other systems and shows the importance of appropriate column-oriented partitioning. Section \ref{sec:relatedwork} discusses related works. Finally, Section \ref{sec:conclusion} concludes this paper.

\section{Push with Column-Oriented Partitions}
\label{sec:abstraction}

The combination of push computation and two-level column-oriented partitions is the guiding idea of \SYS design. The choice enables various efficient optimizations introduced in Section \ref{sec:coreidea}. Without the column-oriented partitioning, fully-out-of-core processing will be impractical even given all other optimizations.

\subsection{Push vs. Pull}

Two basic modes of vertex-centric processing are push and pull modes.
In push mode, each \emph{active} vertex tries to update neighboring vertices through outgoing edges, introducing synchronization overhead.
In pull mode, each vertex updates its state by collecting information from neighbors through incoming edges, avoiding write contentions.
Some systems support both modes and choose one for each iteration, depending on the density of active vertices.

We argue that, while it is essential for in-memory systems to carefully choose the trade-off between excessive synchronization (push) and amount of work (pull), things are different in the out-of-core scenario.
Supporting both modes doubles the working set, consumes more space, and makes caches less helpful.
More importantly, excessive external memory accesses in pull mode would be much more expensive than synchronizations in push mode.

To this end, \SYS opts to keep only the push mode, which naturally enables selective scheduling and avoids unnecessary I/O.
Nevertheless, directly adopting push mode is not enough to enable efficient fully-out-of-core processing. The random writes to external memory could make performance drop dramatically and wear out SSDs in a short time.
\SYS proposes a two-level partitioning strategy to accommodate this issue, which will be introduced next.

\subsection{Two-Level Column-Oriented Partitioning}
\label{sec:partitioning}
Partitioning data across nodes natural for distributed systems, \SYS further adopts a column-oriented partitioning inside each machine, to reduce random access to SSDs and support fully-out-of-core processing.
We will use the terms ``partition'' and ``batch'' to distinguish inter-node and intra-node cases in the following descriptions.


\noindent\textbf{Inter-node partitioning}: \SYS puts vertices with continuous numeric IDs into the same partition\footnote{According to the natural locality of graph data \cite{BoVWFI}, a vertex and its neighbor vertices are more likely to have close IDs. By putting vertices with continuous IDs into the same partition, locality of this kind of graphs could effectively be preserved. Even if the vertices are randomly shuffled, this partition does not make things worse.}. Given a graph and $P$ machines, the vertex set $V$ is partitioned into $P$ disjoint ranges $V_1$, $V_2$, ..., $V_P$. Let $E_i^\text{i}$ and $E_i^\text{o}$ denote the incoming and outgoing edge sets of $V_i$. \SYS tries to make $(\alpha \times |V_i| + |E_i^\text{i}| + |E_i^\text{o}|)$ for each partition $i$ as close as possible, to balance the number of vertices, incoming and outgoing edges of each partition. $\alpha$ is configurable and defaults to $2P - 1$. The partitioning method corresponds to the estimated amount of disk I/O and network traffic per node in Section \ref{sec:internodemessage}.

\begin{figure}[tb!]
    \centering
    \begin{subfigure}[b]{\linewidth}
        \centering
        \includegraphics[scale=.4]{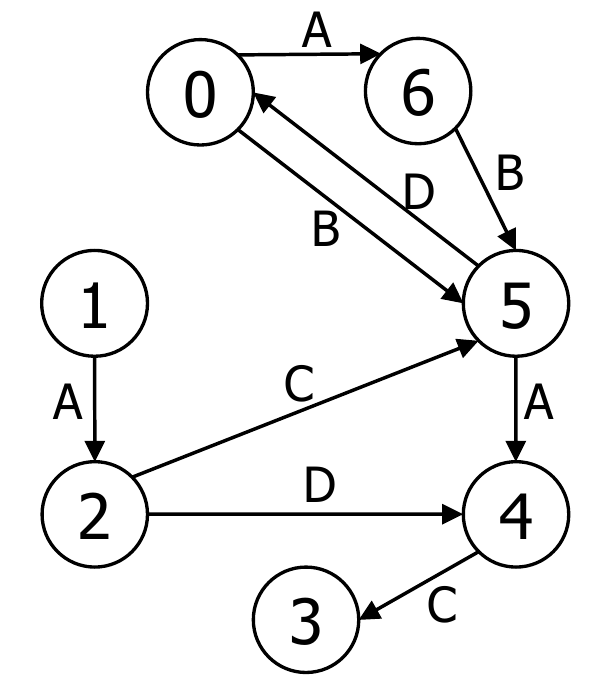}
        \caption{A graph of seven vertices and nine edges. Edge data is a letter.}
        \label{fig:dcsr1}
    \end{subfigure}
    
    \begin{subfigure}[b]{\linewidth}
        \centering
        \includegraphics[width=.85\linewidth]{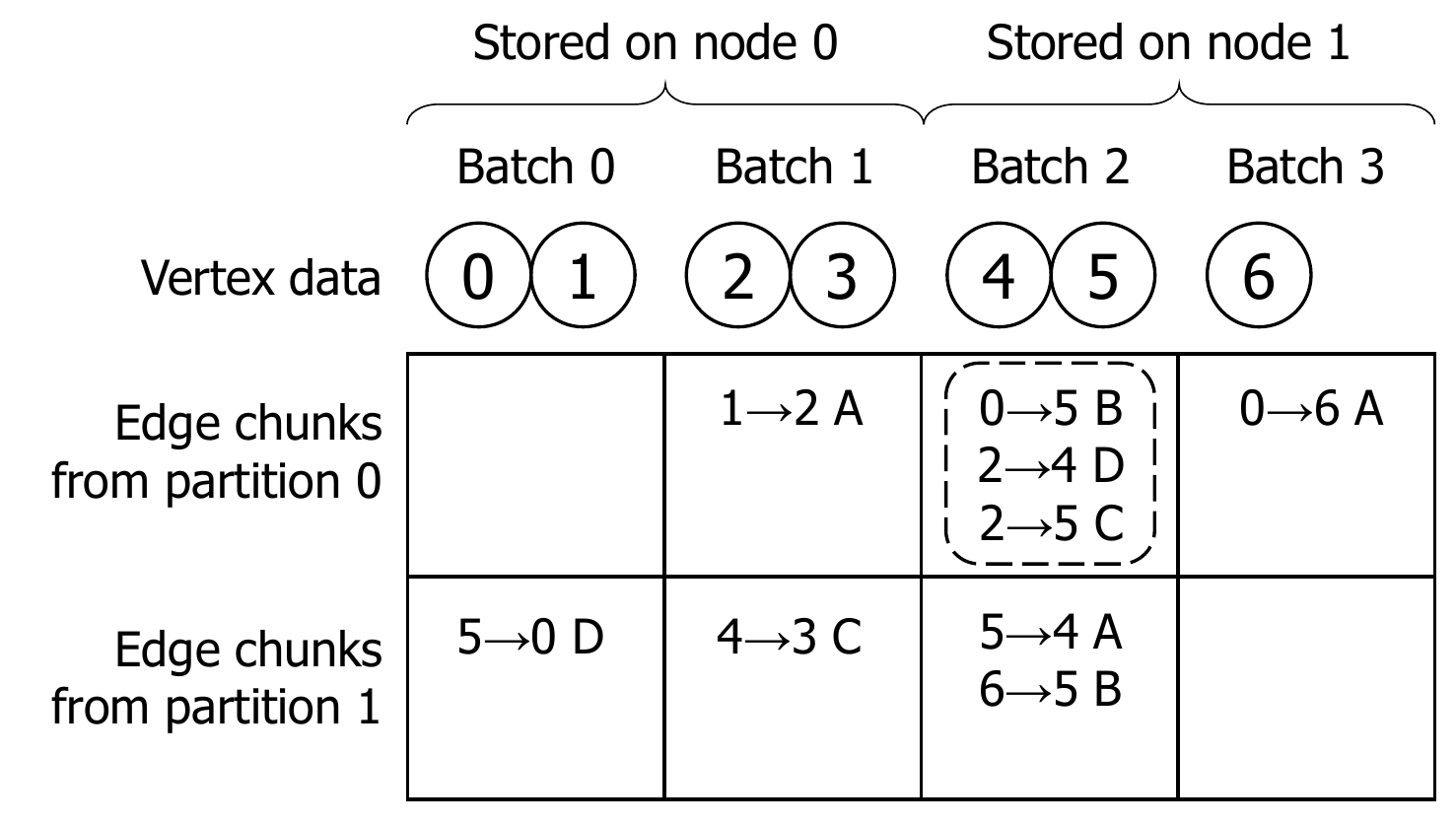}
        \caption{Vertex and edge storage on two nodes. Vertex batch size is 2.}
        \label{fig:dcsr2}
    \end{subfigure}
    
    \begin{subfigure}[b]{.48\linewidth}
        \centering
        \includegraphics[scale=.4]{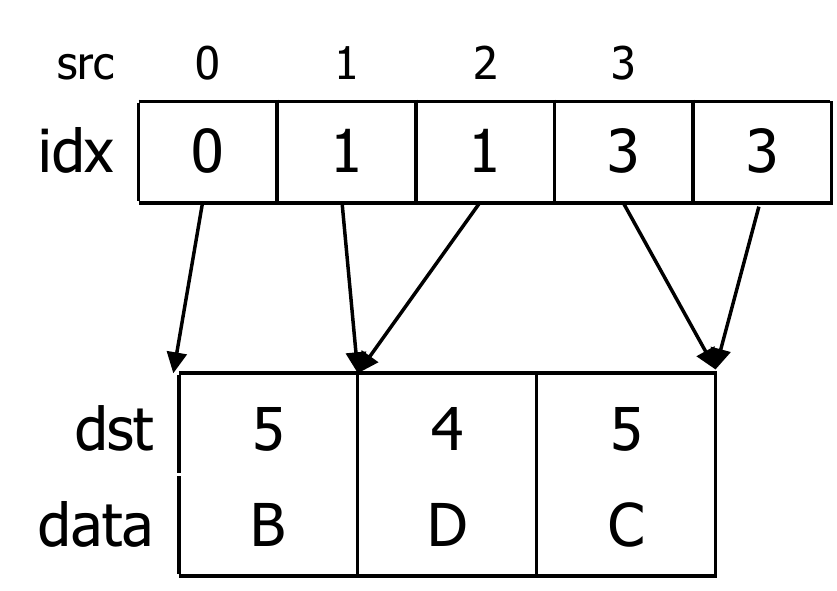}
        \caption{CSR of circled chunk in (b)}
        \label{fig:dcsr3}
    \end{subfigure}
    \hspace{.02\linewidth}
    \begin{subfigure}[b]{.48\linewidth}
        \centering
        \includegraphics[scale=.4]{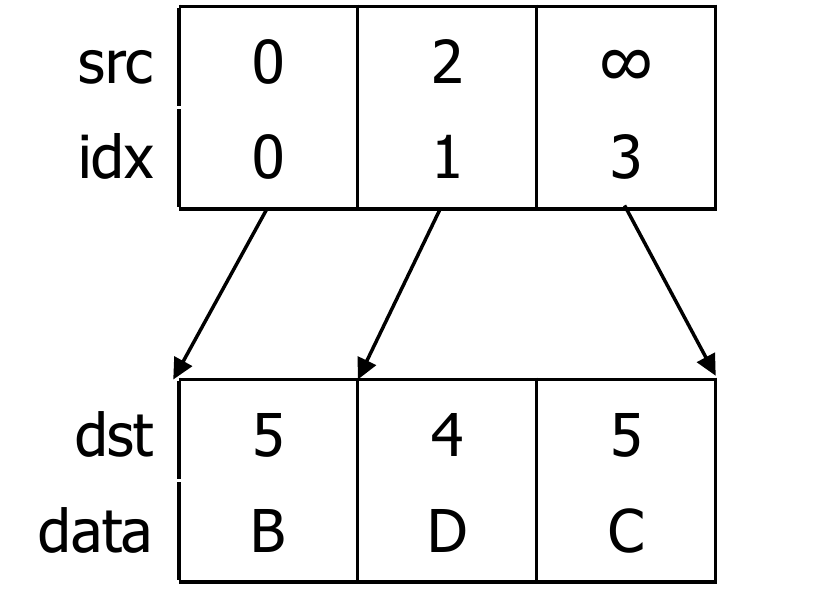}
        \caption{DCSR of circled chunk in (b)}
        \label{fig:dcsr4}
    \end{subfigure}
    \begin{subfigure}[b]{\linewidth}
        \centering
        \includegraphics[width=.8\linewidth]{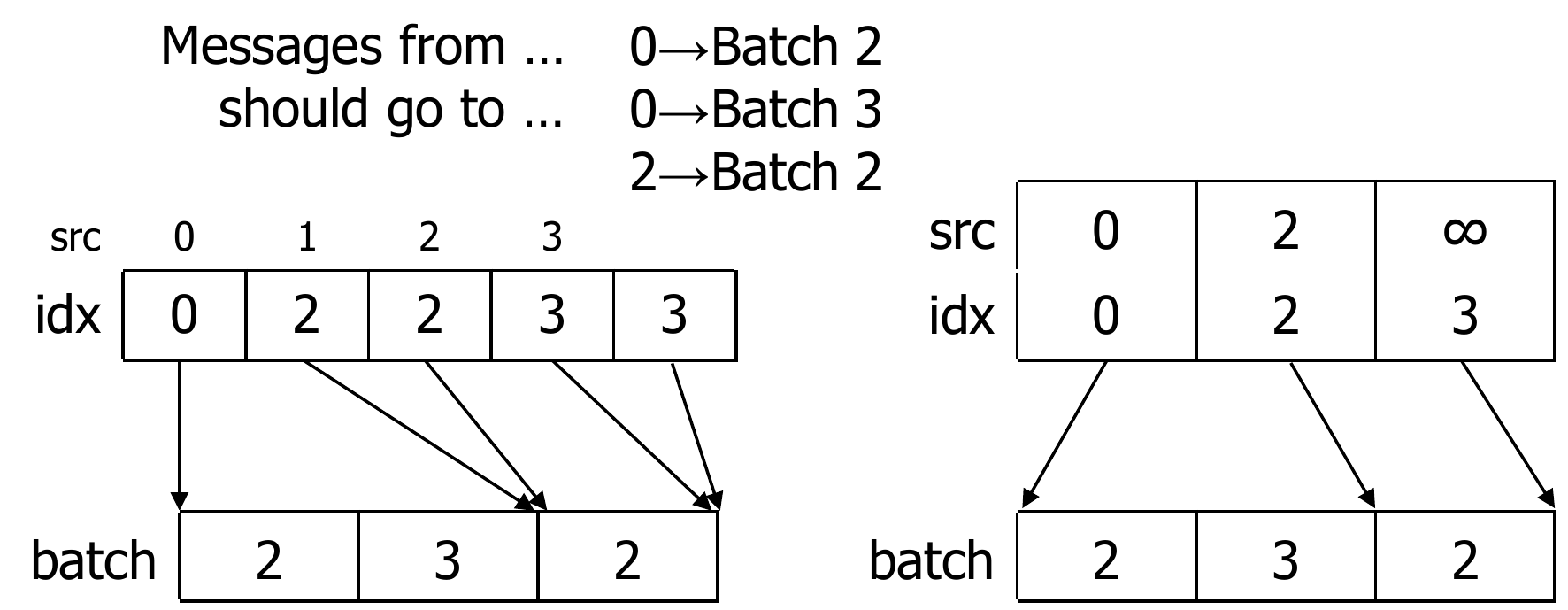}
        \caption{Edges, CSR, and DCSR of dispatching graph from node 0 to 1.}
        \label{fig:dcsr5}
    \end{subfigure}
    \caption{Data storage and representation of a graph.}
\end{figure}

\noindent\textbf{Intra-node batching}: In each machine, \SYS further splits vertices to batches of fixed size\footnote{The last batch may contain fewer vertices.}. Random access to vertex data is limited to the span of one batch, which is critical for fully-out-of-core cases (Section \ref{sec:expbatching} will show the importance of narrowing the random access span by experiments). Figure \ref{fig:dcsr1} shows an example graph of seven vertices and nine edges. Assuming we partition the graph across two nodes, and inside a machine, every two vertices form a batch, as Figure \ref{fig:dcsr2} shows, each machine stores its vertices along with incoming edges. Edges are grouped by the partition of source vertex and the batch of destination, regarded as column-oriented graph partitioning. Each group is stored in an edge chunk where the edges share the same source partition and destination batch.

Vertex batch size is an important parameter and should be wisely chosen. Typically, each partition should contain at least $T$ batches, the number of CPU threads per node. Smaller vertex batches can narrow the span of random vertex data access and help load balancing of multiple threads inside each node since each batch's amount of work differs. However, smaller batches result in more management costs, including metadata storage, and increase the amount of data on disk because edges from the same source vertex are less likely to have destination vertices lying in the same batch, making compressed representations harder. By default, we choose the batch size to be as large as possible, either limited by the memory amount (fully-out-of-core) or by the requirement of load balancing (semi-out-of-core). In fully-out-of-core processing, the size is chosen that vertex data of each batch multiplied by $T$ is less than half of total memory. For the semi-out-of-core case, the size is chosen by experience that each partition contains at least $1.5T$ batches.

\section{\SYS Implementation}

\label{sec:implementation}

\begin{figure}[tb!]
    \centering
    \begin{subfigure}[b]{\linewidth}
        \centering
        \includegraphics[width=.9\linewidth]{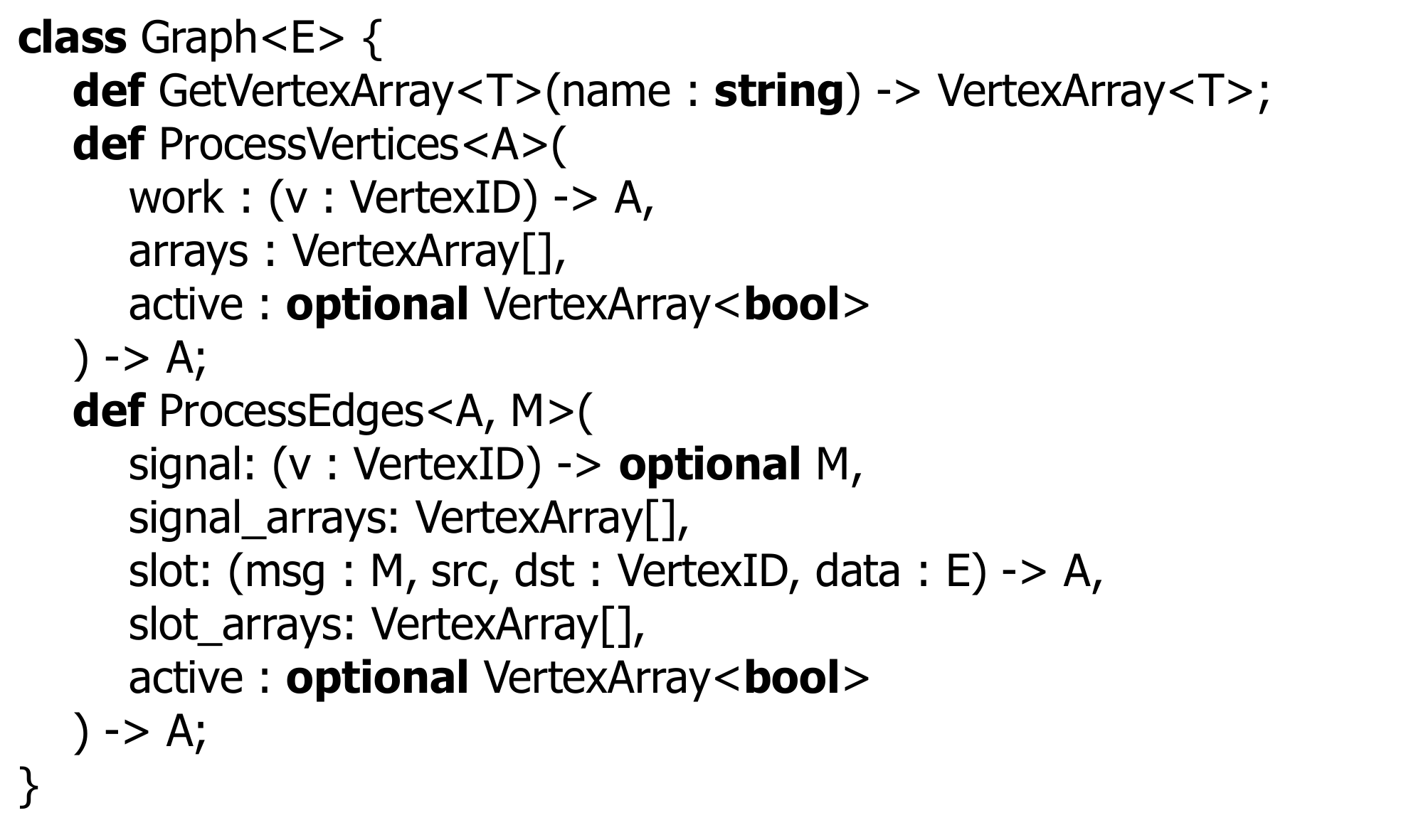}
        \caption{Definitions to \SYS APIs}
        \label{fig:api1}
    \end{subfigure}
    
    \begin{subfigure}[b]{\linewidth}
        \centering
        \includegraphics[width=.9\linewidth]{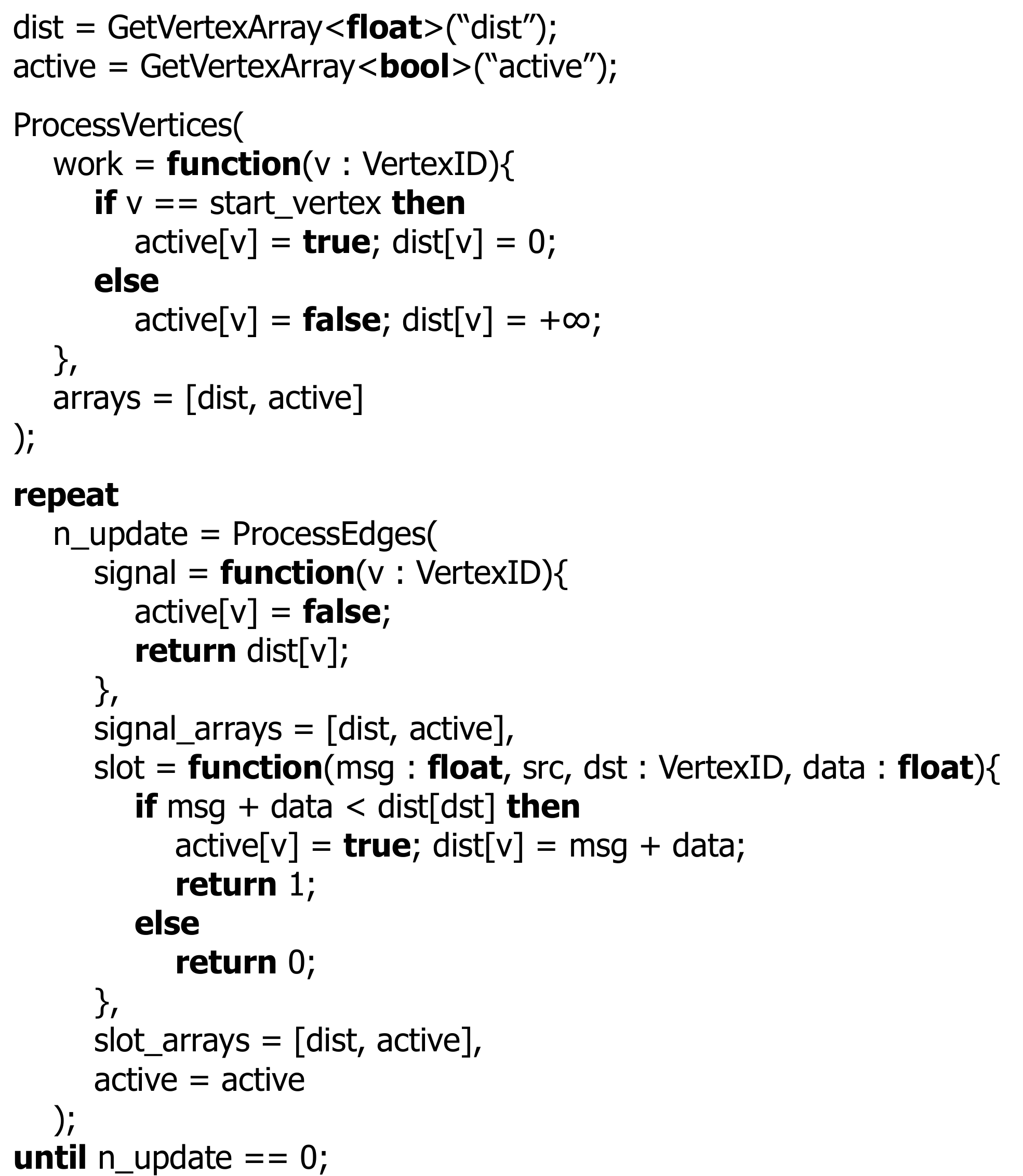}
        \caption{SSSP algorithm based on \SYS APIs}
        \label{fig:api2}
    \end{subfigure}
    \caption{\SYS APIs}
\end{figure}

\SYS mainly provides three APIs: \getdarray defines vertex data, \processVertices computes on vertices, and \processEdges computes for each edge. Typically, each iteration of an algorithm performs one call to \processEdges and zero or a few calls to \processVertices. Figure \ref{fig:api1} gives their pseudo-code definitions, details as follows:

\texttt{GetVertexArray<T>}
creates or loads vertex data of type \texttt{T}, returning a \darray object, which should be passed to \process functions for accesses.

\processVertices
does computations on vertices. \emph{work} is a user-defined function (UDF) called as \emph{work}(\emph{v}), performing user operations on vertex \emph{v}. Return values of \emph{work} are summed and returned by \processVertices. \emph{arrays} provide \darrays that may be accessed by \emph{work}. Users may access data of vertex \emph{v} inside a call to \emph{work}(\emph{v}). If the user supplies \emph{active}, a boolean \darray representing an active set, \emph{work} will only be called for vertices in it.

\processEdges
does computations on edges, following a simple signal-slot model.
Active vertices generate messages, passed through outgoing edges, received and processed by destination vertices\footnote{In case an algorithm needs messages to go ``reversely'' (through incoming edges), the user could call \processEdges on the graph with reversed edges. \darrays can be shared by the original and the reversed graphs. Thus, it is possible to implement any algorithm on graphs with \SYS API, though for a specific application there could be more efficient abstractions.}.
\emph{signal} is a UDF called as \emph{signal}(\emph{src}) and optionally generates (returns) a message from vertex \emph{src}. Data of \emph{src} in \emph{signal_arrays} may be accessed inside a call to \emph{signal}(\emph{src}). If the user supplies \emph{active}, \emph{signal} will only be called for active vertices. 
\emph{slot} is a UDF called as \emph{slot}(\emph{msg}, \emph{src}, \emph{dst}, \emph{data}) if vertex \emph{src} generated a message \emph{msg}, and an edge exists from \emph{src} to \emph{dst} with \emph{data}. Return values of \emph{slot} calls are summed and returned by \processEdges. Data of \emph{dst} in \emph{slot_arrays} may be accessed inside the \emph{slot}(\emph{msg}, \emph{src}, \emph{dst}) call, while data of \emph{src} may not. \SYS guarantees any \emph{signal} call happens before any \emph{slot} call, and \emph{slot} calls do not have data race when accessing \emph{dst} in \emph{slot_arrays}. Thus, no atomic operation is needed when accessing vertex data, which makes it easier for programming.

Figure \ref{fig:api2} demonstrates the use of APIs to implement an SSSP (single-source shortest path) algorithm.
It creates \darrays, initializes them, and repeatedly updates the shortest path until no result changes in the current iteration. As \SYS guarantees, users need not care about data contention for \darrays.

\subsection{Phases of Communication}
\begin{figure}[tb!]
    \centering
    \includegraphics[width=.9\linewidth]{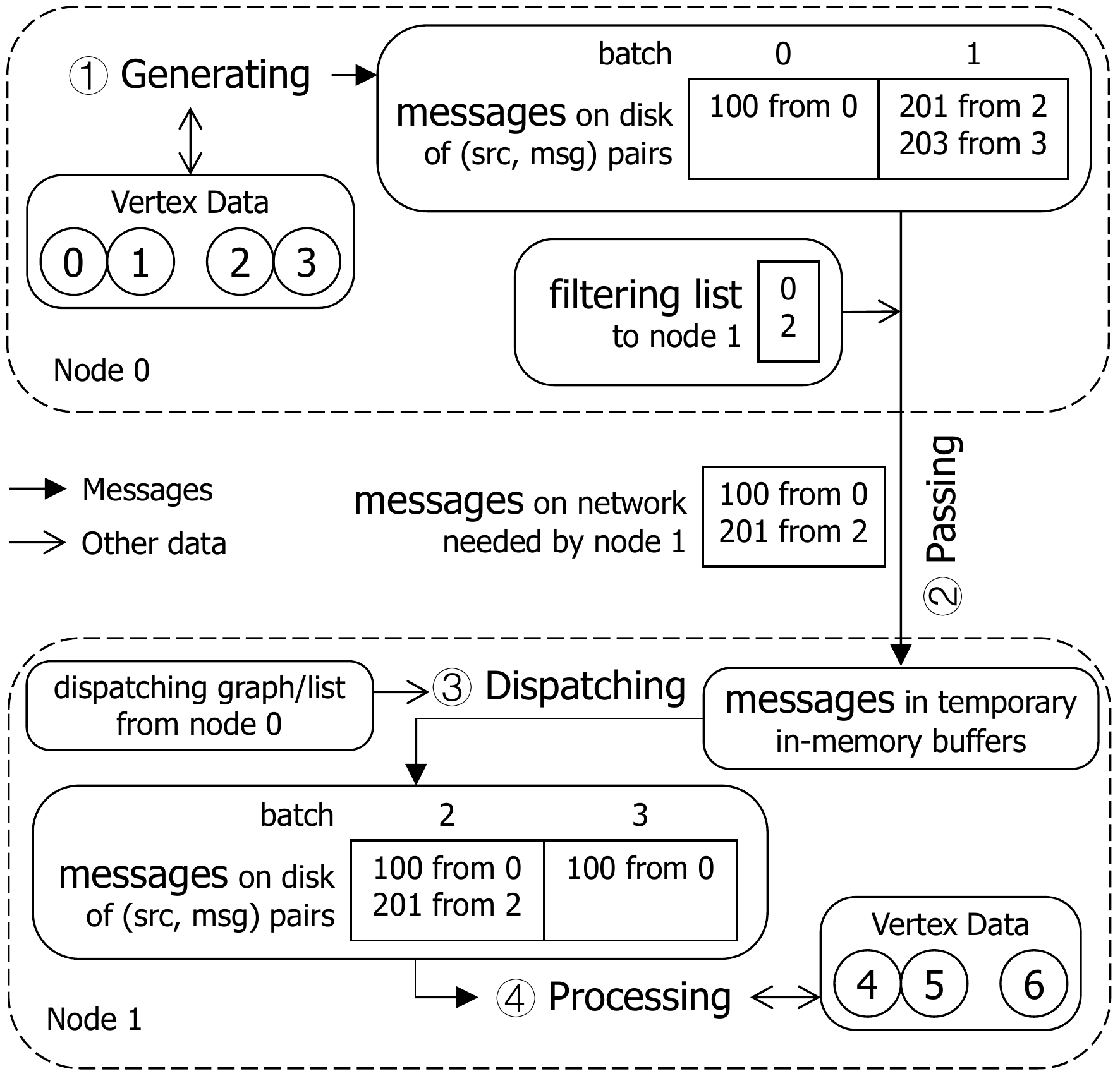}
    \caption{Data flow of \processEdges from node 0 to 1.}
    \label{fig:msg}
\end{figure}

DFOGraph passes each message inter-node and then dispatched it intra-node to batches that need it, which corresponds to the two-level column-oriented partitioning. Our design of \processEdges contains four phases -- generating, inter-node passing, intra-node dispatching, and processing. Figure \ref{fig:msg} shows the message and data flow of communication from node 0 to node 1 using the example graph and configuration above. The goal of each phase is listed as follows:

\begin{enumerate}[leftmargin=*,itemsep=0pt]
    \item Generating: Each batch interacts with its vertex data and saves the messages it generates in \emph{signal} to disk.
    \item Inter-node passing: Each node $i$ sends messages to buffers of each node $j$'s memory. \SYS filters the messages and only sends needed ones. A node needs a message if its source vertex has outgoing edges to the partition.
    \item Intra-node dispatching: Each node dispatches the messages to its batches with dispatching graphs or lists, resulting in a file storing needed messages for each batch.
    \item Processing: Each batch does computation in \emph{slot} and interacts with vertex data, based on its edges and messages.
\end{enumerate}
Section \ref{sec:coreidea} will introduce the additional data needed by each phase to optimize the communication.

\subsection{Fault Tolerance}
\label{sec:faulttolerance}

\begin{figure}[tb!]
    \centering
    \includegraphics[width=.9\linewidth]{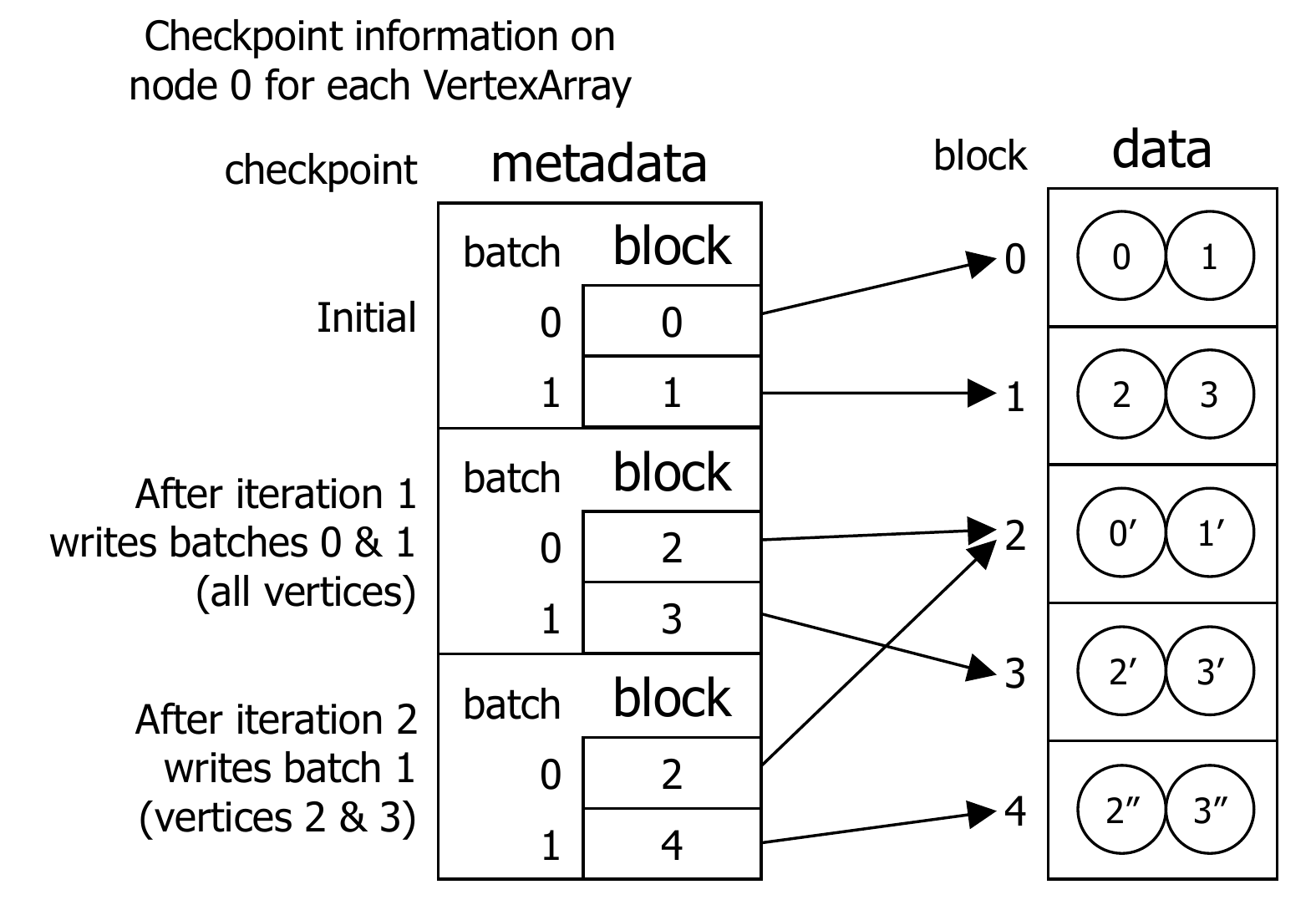}
    \caption{Example checkpoints of a \darray}
    \label{fig:checkpoint}
\end{figure}

In \SYS's computation model, attributes of edges are immutable, while data on vertices can change during the \process functions if UDF writes to \darray. If the user enables checkpointing, \SYS never overwrites data blocks, and redirects all write operations to a new block for each operation. For each \darray, each machine keeps track of the latest data block location of each vertex batch. Thus, \darray structures become persistent.

Figure \ref{fig:checkpoint} shows example checkpoints of a \darray on node 0.
\getdarray creates the initial checkpoint, and each \process call creates a new one. The first call writes both batches on the current machine, while the second call only modifies batch 1 and reuses the data of batch 0 from the previous checkpoint.

With the persistent structure, if \SYS encounters an error during computation, it can recover from the state after the last successful call. Users can configure the number of checkpoints maintained (typically 1 or 2), and \SYS will remove obsolete checkpoints and data blocks by reference counting when a new \process function completes. Thus, \SYS can resume execution after a recoverable failure on any machine, losing progress of no more than one \process call.
The storage overhead includes the metadata and old versions of data, while the computation overhead only includes the metadata as checkpointing does not increase the amount of I/O.

\section{I/O and Communication Optimizations}

\label{sec:coreidea}

Push mode and column-oriented partitions enable multiple optimizations. \SYS uses adaptive CSR and DCSR representations to reduce I/O. It filters messages before sending them to cut back network traffic and develops adaptive strategies for dispatching to balance CPU, disk, and network usage. \SYS further decomposes and pipelines operations related to disk or network, so that it overlaps computation with I/O and communication, and hides the extra latency of the throughput optimizations.
\subsection{Adaptive CSR and DCSR Representations}
\label{sec:storage}

Because of intra-node column-oriented partitioning, edge chunks might be sparse, containing fewer edges than vertices, which calls for the need to use CSR (Compressed Sparse Row) and DCSR (Doubly-Compressed Sparse Row) for edges.

CSR stores two arrays, $idx$ and ($dst$, $data$) pair. Figure \ref{fig:dcsr3} shows the CSR storing the chunk of 3 edges from partition 0 to batch 2 of the example graph.
Given a source vertex $src$, we can iterate over $dst[i]$ and $data[i]$ for $idx[src]\leq i < idx[src+1]$ to find the destination and data of each edge from $src$.
For each chunk, the length of $idx$ equals the number of nodes in $src$'s partition $|V_{src}|$, and ($dst$, $data$) needs $|E|$ space to store each edge of this chunk.

Different from CSR, DCSR stores ($src$, $idx$) pair rather than $idx$ alone, shown in Figure \ref{fig:dcsr4}. Only vertices with outgoing edge correspond to elements in ($src$, $idx$), which is at most $|E|$. DCSR is especially helpful when the chunk is sparse ($|V_{src}|\gg |E|$). However, it does not support $O(1)$ seeking, and needs extra effort to find the required element ($src$, $idx$), bringing more overhead when we only access a few $src$-s.

\SYS always builds DCSR for every edge chunk, and builds CSR for edge chunks where $|V_{src}|/|E| \leq 32$, which is a configurable parameter named ``CSR inflate ratio.'' 
However, even if CSR is available for a chunk, \SYS may read DCSR instead during computation.

Accesses to edges happen in the message processing phase. If both CSR and DCSR are available for this chunk, \SYS decides which to use based on an estimation of processing cost. Edge scanning costs of CSR and DCSR are the same, while the difference is the seeking cost given source vertices. We model DCSR's cost to be two times of the length of ($src$, $idx$) array, $2 \times |V_{src,~~outdeg\neq 0}|$, and CSR's to be $\min(\gamma\times|M|, |V_{src}|)$ where $|M|$ denotes the number of messages, $\gamma$ is a parameter defaulting to 1024 by experience, and $|V_{src}|$ is the length of CSR's $idx$ array. The estimation assumes we scan the ($src$, $idx$) array of DCSR sequentially, while CSR is either scanned or sought, and the cost of each seek equals scanning $\gamma$ elements. With the hybrid strategy of CSR/DCSR, \SYS is adaptive for both dense (most sources generate messages) and sparse (only a few sources have messages) situations.

This method of choosing between CSR and DCSR also applies when using push mode for intra-node dispatching, accessing the dispatching graph, introduced next.

\subsection{Adaptive Strategies for Message Dispatching}

\SYS develops three strategies for dispatching and chooses the best one for each communication. ``Push'' and ``pull'' here are only for message dispatching inside each node, not to be confused with push/pull computation modes.

\noindent\textbf{Push dispatching}: A single thread scans the messages and appends messages to files of batches, which needs a ``dispatching graph'' to determine the destination batches of each message. Figure \ref{fig:dcsr5} shows the dispatching graph used when node 1 receives messages from node 0. Each edge of the graph represents one relationship like ``messages from vertex X should go to batch Y.'' Dispatching graphs are preprocessed and stored as DCSR, and optionally as CSR if acceptable by CSR inflate ratio, similar to the storage of edge chunks. Only after the whole scan finishes can \emph{any} vertex batch enter the processing phase. Thus, the computation cost of push dispatching is low, but the latency is high.

\noindent\textbf{Pull dispatching}: Each batch scans the messages and appends needed ones to its file. Calculated during preprocessing, \SYS uses a list of $src$ vertices needed from each partition to each batch, to perform the pull dispatching. Unlike pushing, once a batch finishes pulling, the message processing phase of the batch can start. The computation cost of pull dispatching is high, but the latency is low for the first batches. \SYS uses pull dispatching if the CPU is idle waiting for messages to reduce the latency, probably when a node is dispatching messages from itself, since messages from other nodes have not arrived yet, and the next phase is idle waiting for the first messages to process.

\noindent\textbf{No dispatching}: Let batches directly read from the messages without dispatching, used when the cost of push/pull dispatching is much higher than the messages themselves.

With hybrid strategies for message dispatching, \SYS can adaptively reduce the processing latency when the CPU load is low and when the number of messages is small.

\subsection{Message Filtering in Inter-Node Passing}

When passing messages from node $i$ to $j$, ``Filtering'' means eliminating the messages that node $j$ does not need, i.e., messages whose \emph{src} does not have outgoing edges to partition $j$. In the example above, vertex 3 generates a message but has no edge to partition 1. Thus it will not appear in the messages from node 0 to node 1.

Calculated in preprocessing, \SYS uses a list of vertices needed from node $i$ to $j$, stored on node $i$, to perform the filtering. The process resembles merging two sorted arrays, and its cost from node $i$ to $j$ equals the number of messages $|M_i|$ plus the length of the list $|L_{ij}|$. If $|L_{ij}|/|M_i|\geq 2$, a configurable threshold, the messages are sent without filtering, to avoid high overhead. As the cost is still proportional to the messages, efficient pipelining is adopted to avoid the doubled latency, introduced next.

\subsection{Careful and Efficient Pipelining}

\SYS carefully decomposes operations related to disk and network, which hides the latency of seeming-expensive steps such as message filtering. Thus, \SYS can efficiently overlap computation, I/O, and communication.

\noindent \textbf{Communication phases}: The phases of \processEdges are pipelined rather than serialized. After a batch of messages completes one phase, it enters the next phase as long as CPU, network, and disk resources are available.

\noindent \textbf{Vertex-parallel jobs}: In \processVertices, batches do not interfere with each other. Thus, they can be processed in parallel. For each batch, \SYS loads $active$ from disk, finishes if no vertex in the batch is $active$, loads $arrays$, calls $work$ with vertices of the batch, and finally writes back dirty data. \SYS overlaps I/O and computation by pipelining the operations. There is overhead to load $arrays$ after scanning $active$ if the $arrays$ are finally needed. Pipelining can mostly hide the overhead, reducing the extra latency to scanning a single batch of $active$ in memory, which is acceptable. Other vertex-parallel situations, such as message generating and processing phases, are similarly pipelined.

\noindent \textbf{Inter-node message passing}: When node $i$ passes messages to node $j$, we pipeline these operations on node $i$: loading messages from disk, filtering them, and sending them to node $j$. Also, \SYS adopts round-robin scheduling: node $i$ sends messages to nodes $i+1, \dots, P-1, 0, \dots, i-1$ in order. Sending to different nodes can be parallel because no data race exists, which happens if extra network bandwidth is available (e.g., the number of generated messages is small) to reduce the latency.

\subsection{Discussions Based on Optimizations}
\label{sec:internodemessage}

\begin{table}[tb!]
  \centering
    \begin{tabular}{c|c|c}
      \toprule[1.5pt]
      Phase & Disk & Network \\
      \midrule[1pt]
      Generate
      & \makecell{Read \& Write $\leq|V_i|$}
      & -- \\
      \midrule[.5pt]
      Pass
      & \makecell{Read $\leq$ \\ $(P-1) \times|V_i|+|E_i^\text{o}|$}
      & \makecell{Send $\leq|E_i^\text{o}|$ \\ ($\leq|V_i|$ to each node)} \\
      \midrule[.5pt]
      Dispatch
      & \makecell{Read \& Write $\leq |E_i^\text{i}|$}
      & \makecell{Receive $\leq|E_i^\text{i}|$ \\ ($\leq|V_j|$ from node $j$)} \\
      \midrule[.5pt]
      Process
      & \makecell{Read $\leq P\times |V_i|+|E_i^\text{i}|$ \\ Write $\leq P \times |V_i|$}
      & -- \\
      \bottomrule[1.5pt]
    \end{tabular}
  \caption{I/O and Communication amount in each phase of \processEdges API on node $i$.}
  \label{tbl:ionetwork}
\end{table}

\begin{table}[tb!]
  \centering
    \begin{tabular}{crrr}
      \toprule[1.5pt]
      Graph & {$|V|$ / Million} & {$|E|$ / Billion} & {Size / GB} \\
      \midrule[1pt]
      {\twi} & 41.7 & 1.47 & 10.9 \\
      {\uk} & 787.8 & 47.61 & 354.7 \\
      {\rmat} & 4 295.0 & 68.72 & 1 024.0 \\
      {\kron} & 274 877.9 & 1 099.51 & 16 384.0 \\
      \bottomrule[1.5pt]
    \end{tabular}
  \caption{Graph datasets for experiments -- size calculated as (source, destination) pair in binary formats of each edge.}
  \label{tbl:datasets}
\end{table}

Now we discuss the features and assumptions of \SYS.

\noindent \textbf{Inter-node partitioning strategy}: We choose the inter-node partitioning method to be $(\alpha \times |V_i| + |E_i^\text{i}| + |E_i^\text{o}|)$ according to an estimation of work on each node, shown in Table \ref{tbl:ionetwork}. ``$\leq$'' indicates a worst-case estimation. Not every vertex produces a message, which relaxes terms like $\leq|V|$. Also, edges from the same source to the same partition reduce the number of messages to send. Thus, terms like $\leq|E|$ are loose in passing and dispatching phases. We compute the amount assuming we use DCSR representation and push dispatching, and filtering is enabled. If \SYS uses another strategy, the amount may increase, but \SYS only does so when the end-to-end time is estimated to be shorter. Total work on each node is approximately $(2P-1)\times |V_i| + |E_i^\text{i}| + |E_i^\text{o}|$, which corresponds to the partitioning method.

\noindent \textbf{Bandwidth assumption}: Any message sent inter-node is useful and corresponds to at least one edge access on the destination node. A node can simultaneously send/receive messages from/to only one peer node at a time (communication with more peers only happens given extra bandwidth). Thus, \SYS only assumes the network bandwidth is comparable to or even slightly less than the disk bandwidth per node. On a slower network, \SYS could operate, but the performance will mainly depend on the network.

\noindent \textbf{Data contention}: All the messages and edges are streamed rather than entirely read in once because \SYS does not assume any of them (even edges from a single vertex) can fit in memory. As the result of the message sending order, any vertex batch of node $i$ processes messages received from nodes $i-1$, \dots, 0, $P-1$, \dots, $i+1$ in order. Each batch processes incoming messages with a single thread, and calculation on messages from one source node happens before that from the next source. Thus, users need not handle data contention or use atomic operation within the $slot$ function.





\section{Experiments}
\label{sec:evaluation}

We implemented \SYS in \char`\~ 3,000 lines of C++ code, with MPI and pthread parallel libraries. The code is compiled by GNU C++ Compiler 9.3.0 with Open MPI 4.0.3 on x86-64 Linux.
Vertex batch size is an essential parameter for the performance. We did not tune it for each experiment but directly chose one according to discussions in Section \ref{sec:partitioning}.

We conducted the experiments on AWS EC2 i3en.3xlarge instances, each equipped with 12 threads of Intel Xeon Platinum 8175M (2 threads per core, base frequency 2.50 GHz), 93.2 GB RAM, 25 Gbps network (measured by \emph{iperf} \cite{iperf}), NVMe SSD (2 GB/s sequential read bandwidth by \emph{fio} \cite{fio}).

We did not artificially limit the memory usage by default. For systems that need a parameter of memory budget (Chaos and GridGraph, as well as HybridGraph for Java heap size), we supplied 88 GB to avoid out-of-memory errors.

\subsection{Graph Algorithms and datasets}

We performed four algorithms in the experiments: PageRank (PR) \cite{page1999pagerank}, Breadth-First Search (BFS), Weak Connecting Component (WCC), and Single-Source Shortest Path (SSSP). In PR, each iteration scans the whole graph, and we perform five iterations in each run. In contrast, during BFS, the number of iterations equals the longest distance from the starting vertex, and each edge is only scanned once within a run. Patterns of WCC and SSSP are between PR and BFS, where each iteration may not scan the whole graph, and an edge is likely to be accessed multiple times in each run.

We used four graph datasets, shown in Table \ref{tbl:datasets}. \twi social graph
\cite{kwak2010twitter} and \uk web graph \cite{BMSB} (downloaded from \cite{BRSLLP,BoVWFI}) are from the real world, while the others are synthetic with R-MAT \cite{chakrabarti2004r} and Kronecker \cite{leskovec2005realistic} (large real-world graphs mentioned \cite{backstrom2012four, ShenTu} are currently not published).
\uk has a large diameter and needs \char`\~2500 iterations for algorithms except for PR. Some systems failed to finish them in time due to high overhead, given a small active set.

\subsection{Experiments on Single Machine}

\begin{table}[tb!]
  \centering
    \begin{tabular}{ccccc}
      \toprule[1.5pt]
      \multicolumn{2}{c}{Workload} & \SYS & GridGraph & FlashGraph \\
      \midrule[1pt]
      \parbox[t]{0mm}{\multirow{5}{*}{\rotatebox[origin=c]{90}{\twi}}}
      & Prep & 31.99 & 62.75 & 618.29 \\
      & PR   & 46.77 & 34.18 &    D   \\
      & BFS  &  8.60 &  9.94 &  10.33 \\
      & WCC  & 42.48 & 10.38 &    D   \\
      & SSSP & 48.11 & 29.46 &    D   \\
      \hline
      \parbox[t]{0mm}{\multirow{5}{*}{\rotatebox[origin=c]{90}{\uk}}}
      & Prep & 1508 &   3178 & M \\ 
      & PR   &  804 &   1569 &   1235* \\
      & BFS  &  870 & >43200 &    556* \\
      & WCC  & 3590 & >43200 &      D* \\
      & SSSP & 3906 & >43200 &      D* \\
      \hline
      \multicolumn{3}{c}{Relative time}
      & >2.52$\times$ & 1.06$\times$ \\
      \bottomrule[1.5pt]
    \end{tabular}
  \caption{Performance comparison among \SYS, GridGraph, and FlashGraph. Time reported in seconds. ``Prep'' reports preprocessing time. ``Relative time'' reports the geometric mean of the relative time of four algorithms compared with \SYS, except for ``D'' values. ~~~~~~~M -- Out of memory; \\
  * -- Using the graph preprocessed elsewhere as input; \\
  D -- No progress for >1 hour, probably in deadlock.
  }
  \label{tbl:singlenoderesults}
\end{table}

\begin{table}[tb!]
  \centering
    \begin{tabular}{cccccc}
      \toprule[1.5pt]
      \multicolumn{2}{c}{Workload} & \makecell{DFO-\\Graph} & Chaos & \makecell{Hybrid-\\Graph} & Gemini \\
      \midrule[1pt]
      \parbox[t]{0mm}{\multirow{5}{*}{\rotatebox[origin=c]{90}{\twi}}}
      & Prep & 12.43 &  61.3 & 498 & 54.1  \\
      & PR   & 10.56 &  45.9 & 116 &  2.59 \\
      & BFS  &  6.95 &  37.5 &  75 &  1.91 \\
      & WCC  & 20.16 & 165.2 & 184 &  4.34 \\
      & SSSP & 20.39 & 244.5 & 268 &  8.97 \\
      \hline
      \parbox[t]{0mm}{\multirow{5}{*}{\rotatebox[origin=c]{90}{\uk}}}
      & Prep & 254 &    564 &  1762   & 1036   \\
      & PR   &  42 &   1664 &  1452   &   14.0 \\
      & BFS  & 861 & >43200 & >2593${}_{\text{R}53}$ &  108.5 \\
      & WCC  & 950 & >43200 & >8180${}_{\text{R}124}$ &   81.9 \\
      & SSSP & 966 & >43200 & 14208   &  155.6 \\
      \hline
      \parbox[t]{0mm}{\multirow{5}{*}{\rotatebox[origin=c]{90}{\rmat}}}
      & Prep & 1105 &   3746   & R* &  M \\
      & PR   &  921 &   4404   & -- & -- \\
      & BFS  &  654 &   5340   & -- & -- \\
      & WCC  & 3611 &  24553${}_\text{C}$ & -- & -- \\
      & SSSP & 4859 & >43200${}_\text{C}$ & -- & -- \\
      \hline
      \parbox[t]{0mm}{\multirow{2}{*}{\rotatebox[origin=c]{90}{\kron}}}
      & Prep & 23428 & \multirow{4}{*}{\makecell{Prep+PR1\\>86400}} & R* &  M \\
      & \makecell{PR1\\(1 iter.\\of PR)}  & 26499 & & -- & -- \\
      \hline
      \multicolumn{3}{c}{Relative time} & >12.94$\times$ & >10.82$\times$ & 0.21$\times$ \\
      \bottomrule[1.5pt]
    \end{tabular}
  \caption{Performance comparison among \SYS, Chaos, HybridGraph, and Gemini on eight nodes. ``Prep'' reports preprocessing or loading time. ``Relative time'' reports the geometric mean of the relative time of four algorithms compared with \SYS.~~~~~~~~M -- Out of memory; \\
  R$i$ -- Crashed after at $i$-th iteration (\char`\~2500 iterations in total); \\
  R* -- Crashed since HybridGraph assumes $|V|<2^{31}$; \\
  C -- Crashed, resumed from checkpoint, reporting total time;
  }
  \label{tbl:distributedresults}
\end{table}

We ran the four algorithms on the two smaller graphs using \SYS, GridGraph \cite{GridGraph}, and FlashGraph \cite{FlashGraph} on a single machine (instance) and recorded time for preprocessing\footnote{Sorting is not included in preprocessing.} and for each algorithm, shown in Table \ref{tbl:singlenoderesults}. \SYS needs input edges in order, but the other two do not. FlashGraph needs text data, but the other two use binary data. Thus, the preprocessing time is not directly comparable. We experienced issues of FlashGraph\footnote{We successfully preprocessed and ran algorithms on a small graph \emph{wiki-Vote}, but encountered two main issues in the experiments: a) FlashGraph failed to preprocess \uk with 93 GB RAM, crashed with ``cannot allocate memory'' messages after sorting the edges. b) FlashGraph reproducibly hangs at the first iteration for some workload, showing no progress for hours, thus having no data to report. We believe it is probably in a deadlock.}, as marked in the table.

GridGraph, with a significant overhead during iterations of a small active set, cannot finish algorithms except PR on \uk. \SYS achieves 1.95$\times$ and 1.54$\times$ speed on \uk PR compared to GridGraph and FlashGraph, respectively, showing the effect of intra-node partitioning for reducing data contention and overlapping computation and I/O. The performance of \SYS in \uk BFS compared to FlashGraph is 0.64$\times$, because \SYS currently suffers from the cost of scanning the whole $active$ bitmap. On the other hand, FlashGraph may store active vertices in a queue-like structure to reduce massive scans on vertices.

\SYS does not have an outstanding performance on \twi, which fits in memory. It suffers from the cost of dispatching messages and managing vertex batches, which cannot be effectively overlapped by I/O when operating in memory. The overhead of the whole bitmap scan makes \SYS perform worse on WCC and SSSP in memory.

Despite the overhead, \SYS achieves >2.52$\times$ and 1.06$\times$ overall speed over GridGraph and FlashGraph, even considering in-memory situations. \SYS on a single node provides performance comparable to systems dedicated to a single machine, especially for out-of-core scenarios.

\subsection{Experiments in Distributed Scenario}

We ran the four algorithms on the three smaller graphs using four systems\footnote{We did not compare with TurboGraph++ because we requested for source code or binary executable but did not get them from the authors.}, including \SYS, Chaos \cite{Chaos}, HybridGraph \cite{HybridGraph}, and Gemini \cite{Gemini} on eight machines (instances), and recorded time for preprocessing or loading and for each algorithm\footnote{Chaos did not separately report loading and computing time for the first iteration and the results are calculated, e.g. 5 PR iterations equals 6 PR iterations (including loading) minus 1 PR iteration (including loading).}, shown in Table \ref{tbl:distributedresults}. For the trillion-edge graph \kron, we only attempted one iteration of PR. \SYS and HybridGraph need sorted input while others do not. HybridGraph is supplied with text files in Hadoop File System (HDFS) \cite{HDFS} while the others use binary input. Thus, the preprocessing/loading time is not directly comparable.

In-memory system Gemini can only process the two smaller graphs. The current code of HybridGraph assumes $|V|<2^{31}$ and cannot process larger ones. It may also crash at different iterations on some workload, and we report the longest run among three attempts. Regarding crashed runs, the speed is calculated with elapsed time before it crashes, which is only unfair to \SYS. Chaos also crashed in some runs and resumed from checkpoints, reporting the total time.

\SYS outperforms Chaos and HybridGraph in all tested cases. Our observations show network is the bottleneck of Chaos, and it did not efficiently utilize the CPU and disks. HybridGraph did not fully utilize either CPU, disk, or network, which may be a source of limited performance. The overall speed of \SYS compared with Chaos and HybridGraph is >12.94$\times$ and >10.82$\times$ on average.

It is not surprising that \SYS is slower than Gemini in all cases it can run on. \SYS suffers from managing batches and additional passes to scan the messages. Also, Gemini can switch between push and pull mode, while \SYS only supports push mode to reduce the I/O overhead. The overall performance of \SYS is 21\% of Gemini.

For \kron, a fully-out-of-core case (2 TB vertex data vs. 746 GB aggregate memory), Chaos did not complete loading and one iteration of PR in 24 hours, while \SYS preprocessed it in 6.51 hours and finished one iteration in 7.36 hours. This graph is more complicated\footnote{We did not use the same graph because it would be semi-out-of-core in the current experiment settings of eight nodes.} than trillion-edge graphs used by experiments of Chaos and GraM papers because it contains 4$\times$ of vertices.

\subsection{Importance of Intra-Node Batching}

\label{sec:expbatching}

To evaluate the effect of intra-node batching to narrow the random accessing span, especially for fully-out-of-core situations, we ran one PR iteration on \emph{KRON-34} (17.2 billion vertices, 68.7 billion edges) with four nodes. The memory is enough or insufficient (extra memory locked in another sleeping process) to store vertex data.

We compare \SYS with a modified version disabling intra-node batching. There are fewer edge chunks without the batches, each storing edges from the same source partition and to the same destination partition (rather than destination batch). No message dispatching happens, and threads of each node process messages in parallel. \darrays are memory-mapped and need atomic operations.

\begin{table}[tb!]
  \centering
    \begin{tabular}{cccc}
      \toprule[1.5pt]
      \makecell{Memory\\per node} & No batching & Batching & \makecell{Batching\\speed}\\
      \midrule[1pt]
      24 GB & >21600 & 1395 & > $15.48\times$ \\
      93.2 GB & 1232 & 1337 & $0.92\times$ \\
      \bottomrule[1.5pt]
    \end{tabular}
  \caption{Time to perform one iteration of PageRank on \emph{KRON-34} with four nodes. Vertex data is 128 GB.}
  \label{tbl:intranodebatching}
\end{table}

Table \ref{tbl:intranodebatching} shows the results. Given insufficient memory (fully-out-of-core), \SYS without batching suffers from massive page swaps as randomly accessing the disk. It did not finish one iteration in 6 hours, even though the vertex data is only 1/3 larger than aggregate memory (128 GB vs. 96 GB). With batching, \SYS performed one iteration in 23 minutes (>15.48$\times$). With enough memory, the batching method needs additional scans during dispatching, while the non-batching method needs atomic operations. Batching brings only 8\% overhead even in the semi-out-of-core case and enables efficient fully-out-of-core processing.

\subsection{I/O and Communication Efficiency}

\begin{figure}[tb!]
    \begin{subfigure}[b]{\linewidth}
    \centering
    \includegraphics[width=\linewidth]{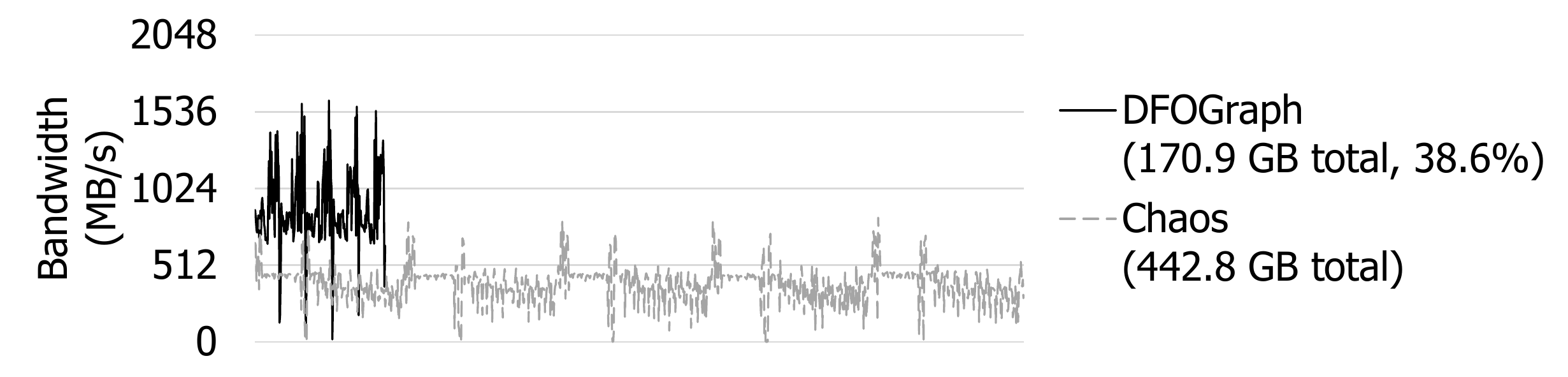}
    \caption{Disk}
    \end{subfigure}
    \begin{subfigure}[b]{\linewidth}
    \centering
    \includegraphics[width=\linewidth]{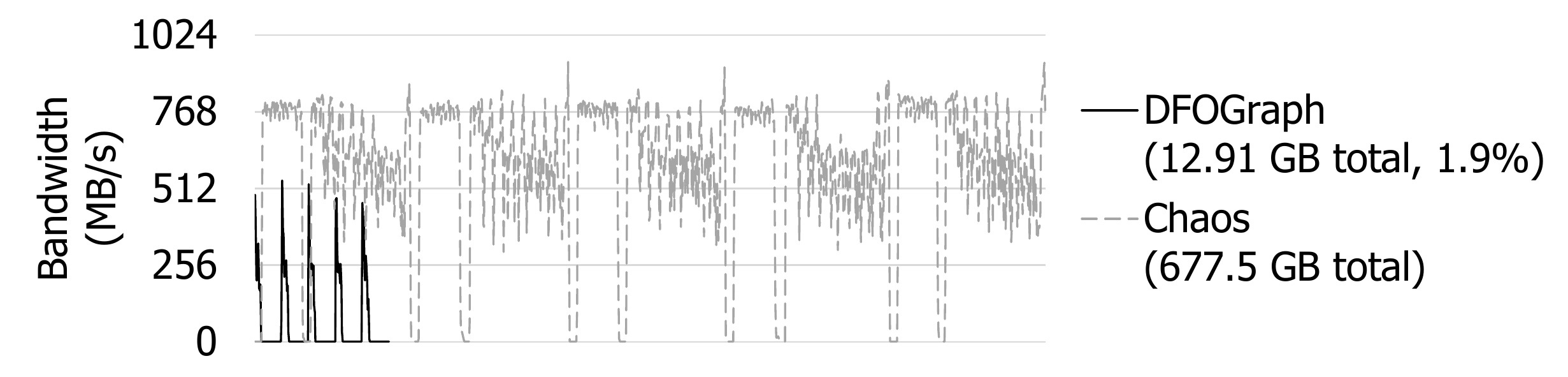}
    \caption{Network}
    \end{subfigure}
    \caption{Traffic over time of \SYS and Chaos, running PageRank on \rmat with 8 nodes.}
    \label{fig:comm}
\end{figure}

\begin{table}[tb!]
  \centering
    \begin{tabular}{cccccc}
      \toprule[1.5pt]
      Algo & $P=1$ & $P=2$ & $P=4$ & $P=8$ & $P=16$ \\
      \midrule[1pt]
      Prep
      & 8230
      & \makecell{3987\\2.06$\times$}
      & \makecell{2480\\3.32$\times$}
      & \makecell{1105\\7.45$\times$}
      & \makecell{ 505\\16.30$\times$}
      \\
      \hline
      PR
      & 7700
      & 5320
      & 2290
      &  921
      &  517
      \\
      BFS
      & 5249
      & 3051
      & 1520
      &  654
      &  209
      \\
      WCC
      & 17055
      & 15246
      &  6789
      &  3611
      &   875
      \\
      SSSP
      & 28436
      & 19400
      & 10057
      &  4859
      &  1003
      \\
      \hline
      \multicolumn{2}{c}{Overall speed}
      & 1.42$\times$
      & 3.01$\times$
      & 6.56$\times$
      & 21.32$\times$
      \\
      \bottomrule[1.5pt]
    \end{tabular}
  \caption{Performance comparison of \SYS using 1, 2, 4, 8 and 16 nodes on \rmat. Time reported in seconds. ``Overall speed'' reports the geometric mean of relative speed of 4 algorithms compared with $P=1$.
  }
  \label{tbl:scaling}
\end{table}

We ran five iterations of PR using \SYS and Chaos on \rmat with eight nodes, recording the disk and network traffic on the first node, shown in Figure \ref{fig:comm}. Chaos did not fully utilize the disk because it generated a massive number of messages and got stuck on it. \SYS issues only 1.9\% messages compared with Chaos, and can better use the disk. Also, \SYS benefits from its adaptive CSR and DCSR representation, reducing the I/O to 38.6\%. With a relaxed assumption on the network, it is possible to run \SYS efficiently on more types of hardware configurations.

\subsection{Impact of Vertex Batch Size Operating Semi-out-of-Core}

\begin{figure}[tb!]
    \centering
    \includegraphics[width=\linewidth]{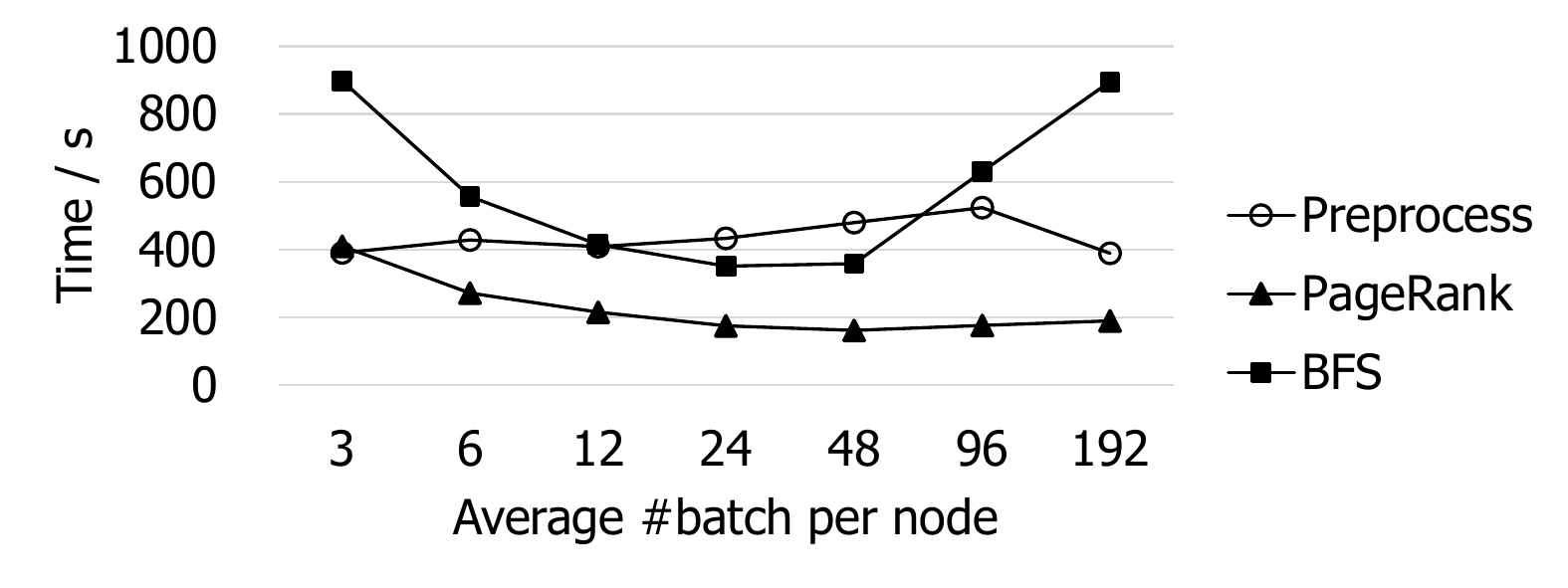}
    \caption{Time comparison choosing different batch size, running PageRank and BFS on \uk with 4 nodes.}
    \label{fig:batchsize}
\end{figure}

Figure \ref{fig:batchsize} shows the running time of preprocessing, PageRank and BFS on \uk with 4 nodes using different batch sizes. As the CPU has $T=12$ threads per node, number of batches less than 12 per node suffers from poor performance. For PageRank and BFS, the optimal choice lies between $2T$ and $4T$. The performance of BFS significantly decreases for much smaller batches because fewer edge chunks are accepted by CSR inflate ratio and more chunks can only be processed with DCSR which needs more efforts. This can also demonstrate the advantage of adaptive CSR and DCSR representation.

\subsection{Scalability}

We ran the four algorithms on \rmat using \SYS on 1, 2, 4, 8, and 16 nodes, the result shown in Table \ref{tbl:scaling}. The overall speedup of 8 nodes is 6.56$\times$, and the speedup of 16 nodes is 21.32$\times$. Also, preprocessing with 8 and 16 nodes can get 7.45$\times$ and 16.30$\times$ speedup, respectively. \SYS sometimes achieves super-linear speedup because the amount of physical memory also increases given more nodes, making some on-disk data more likely appear in the page cache.

\section{Related Work}
\label{sec:relatedwork}

As graph processing becomes a hotspot in academia and industry research, a series of graph processing systems and optimizations have been designed and developed.

\noindent\textbf{Single-Machine Systems}: Ligra \cite{Ligra}, Ligra+ \cite{Ligra+}, and others \cite{Galois,Polymer} are designed in the scenario of single-machine in-memory graph processing. These systems cannot process large graph which cannot fit into memory. A machine can process larger-scale graphs by storing it in external memory. During processing, the graph is read into memory from external memory in chunks and replaced by next chunks. Many systems \cite{GraphChi,TurboGraph,GridGraph,FlashGraph,NXGraph,XStream,Graphene,Mosaic,ai2018clip} are developed under this scenario. X-Stream \cite{XStream} traverses graph by streaming edges, generating updates, and applying the updates. GridGraph \cite{GridGraph} uses a 2-level hierarchical partitioning schema and dual sliding windows to process each data block, which updates vertices on-the-fly and skips unnecessary data blocks. These systems are limited by storage on a single node, thus lack scalability to more massive graphs.

\noindent\textbf{Distributed In-Memory Systems}: Many systems \cite{Pregel,Gemini,Giraph,Hama,GraphLab,PowerGraph,PowerSwitch,PowerLyra,Cyclops,GraphX,xiao2017tux2} lie in this category. Pregel \cite{Pregel} (and open-source versions \cite{Giraph,Hama}) partitions vertices with hashing and process each iteration synchronously with BSP (bulk synchronous processing). GraphLab \cite{GraphLab} process graphs with asynchronous methods. PowerGraph \cite{PowerGraph} proposes GAS (Gather-Apply-Scatter) programming model and master-mirror notion for vertices. Gemini \cite{Gemini} combines push and pull and provides signal-slot API. TuX$^2$ \cite{xiao2017tux2} connects graph models to machine learning. GraphX \cite{GraphX} is built on Spark \cite{Spark}. These systems are costly to process extreme-scale graphs.

\noindent\textbf{Distributed Out-of-Core Systems}: Extending X-Stream \cite{XStream} from a single machine to a cluster, Chaos \cite{Chaos} is a distributed out-of-core system that can process extremely large-scaled graphs. Like X-Stream, Chaos is edge-centric and applies GAS model to each iteration. Targeting similar out-of-core situations, \SYS can outperform Chaos by >12.94$\times$ in experiments, improving the efficiency and scalability of out-of-core processing, and relaxing the assumptions on networks. TurboGraph++ \cite{TurboGraph++} is an extension of a single-machine out-of-core system TurboGraph \cite{TurboGraph}. It uses a balanced buffer-aware partition, nested windowed streaming model (NWSM), and GAS model to achieve scalable and fast graph analysis.  Pregelix \cite{pregelix}, GraphD \cite{graphd}, and HybridGraph \cite{HybridGraph} are Pregel-like distributed semi-out-of-core systems, and it is not trivial to apply their techniques to fully-out-of-core cases.

\noindent\textbf{Hilbert ordering}: COST \cite{mcsherry2015scalability} uses Hilbert ordering to optimize single-threaded graph processing, and \cite{Mosaic} further designs Hilbert-ordered tiles, enabling multi-threaded Hilbert-ordered graph processing. It helps the storage of graphs be more compact and cache-friendly. Though \SYS currently uses CSR and DCSR representations, Hilbert-ordering may be meaningful when the memory limits the vertex batch size, and random accesses likely drop out of CPU cache.

\noindent\textbf{Reducing I/O}: Memory consumption of \SYS mainly includes vertex data buffers and dispatching buffers, and the OS manages the remaining memory for page buffers or caches. If we move this part of memory management into the graph processing system, we may devise wiser cache policies \cite{zhao2019cacheap,lee2019pre}. Besides, dynamically adjusting edge storage structures among iterations for out-of-core processing \cite{vora2016load} is orthogonal and can be applied upon \SYS. Other recent work includes modifying the synchronous neighborhood-scan model to reduce the I/O amount \cite{ai2017squeezing,vora2019lumos}.

\section{Conclusion}
\label{sec:conclusion}


In this paper, we present \SYS, a distributed fully-out-of-core graph processing system. \SYS applies vertex-centric push computation and a two-level column-oriented graph partition strategy to prevent message combining and optimize memory usage and narrow the span of random access to enable fully-out-of-core push processing. \SYS adopts a hybrid graph representation of CSR and DCSR, reducing the data size and enabling fine-grained vertex-centric edge access. \SYS implements adaptive message processing strategies and pipelined computation and disk/network operations to balance CPU, network, and storage. \SYS provides easy-to-use APIs of push computing model by signal and slot functions, and users even need not care about data contention.

The core idea is combining push computation and two-level column-oriented partition. Pushing creates more opportunities for optimizing I/O and communication. Without the intra-node partitioning, fully-out-of-core processing is impractical because of excessive random accesses on disk.

With all these techniques applied, \SYS optimizes the I/O and the communication efficiency, and avoids unnecessary disk and network operations and achieves scalability and capacity. Our experiments show \SYS significantly outperforms other distributed out-of-core systems like Chaos and HybridGraph, and comparable to single-machine out-of-core systems GridGraph and FlashGraph.

\bibliography{bibfile}


\begin{thebibliography}{52}


\ifx \showCODEN    \undefined \def \showCODEN     #1{\unskip}     \fi
\ifx \showDOI      \undefined \def \showDOI       #1{#1}\fi
\ifx \showISBNx    \undefined \def \showISBNx     #1{\unskip}     \fi
\ifx \showISBNxiii \undefined \def \showISBNxiii  #1{\unskip}     \fi
\ifx \showISSN     \undefined \def \showISSN      #1{\unskip}     \fi
\ifx \showLCCN     \undefined \def \showLCCN      #1{\unskip}     \fi
\ifx \shownote     \undefined \def \shownote      #1{#1}          \fi
\ifx \showarticletitle \undefined \def \showarticletitle #1{#1}   \fi
\ifx \showURL      \undefined \def \showURL       {\relax}        \fi
\providecommand\bibfield[2]{#2}
\providecommand\bibinfo[2]{#2}
\providecommand\natexlab[1]{#1}
\providecommand\showeprint[2][]{arXiv:#2}

\bibitem[\protect\citeauthoryear{??}{fio}{[n.d.]}]%
        {fio}
 \bibinfo{year}{[n.d.]}\natexlab{}.
\newblock \showarticletitle{Fio - Flexible {I/O} Tester}.
\newblock \bibinfo{journal}{\emph{https://github.com/axboe/fio}}
  (\bibinfo{year}{[n.\,d.]}).
\newblock


\bibitem[\protect\citeauthoryear{??}{Gir}{[n.d.]}]%
        {Giraph}
 \bibinfo{year}{[n.d.]}\natexlab{}.
\newblock \showarticletitle{Giraph}.
\newblock \bibinfo{journal}{\emph{http://giraph.apache.org}}
  (\bibinfo{year}{[n.\,d.]}).
\newblock


\bibitem[\protect\citeauthoryear{??}{Ham}{[n.d.]}]%
        {Hama}
 \bibinfo{year}{[n.d.]}\natexlab{}.
\newblock \showarticletitle{Hama}.
\newblock \bibinfo{journal}{\emph{https://hama.apache.org}}
  (\bibinfo{year}{[n.\,d.]}).
\newblock


\bibitem[\protect\citeauthoryear{??}{ipe}{[n.d.]}]%
        {iperf}
 \bibinfo{year}{[n.d.]}\natexlab{}.
\newblock \showarticletitle{iPerf - The ultimate speed test tool for {TCP},
  {UDP} and {SCTP}}.
\newblock \bibinfo{journal}{\emph{https://iperf.fr}}
  (\bibinfo{year}{[n.\,d.]}).
\newblock


\bibitem[\protect\citeauthoryear{Ai, Zhang, Wu, Qian, Chen, and Zheng}{Ai
  et~al\mbox{.}}{2017}]%
        {ai2017squeezing}
\bibfield{author}{\bibinfo{person}{Zhiyuan Ai}, \bibinfo{person}{Mingxing
  Zhang}, \bibinfo{person}{Yongwei Wu}, \bibinfo{person}{Xuehai Qian},
  \bibinfo{person}{Kang Chen}, {and} \bibinfo{person}{Weimin Zheng}.}
  \bibinfo{year}{2017}\natexlab{}.
\newblock \showarticletitle{Squeezing out all the value of loaded data: An
  out-of-core graph processing system with reduced disk {I/O}}. In
  \bibinfo{booktitle}{\emph{2017 {USENIX} Annual Technical Conference ({USENIX}
  {ATC} 17)}}. \bibinfo{pages}{125--137}.
\newblock


\bibitem[\protect\citeauthoryear{Ai, Zhang, Wu, Qian, Chen, and Zheng}{Ai
  et~al\mbox{.}}{2018}]%
        {ai2018clip}
\bibfield{author}{\bibinfo{person}{Zhiyuan Ai}, \bibinfo{person}{Mingxing
  Zhang}, \bibinfo{person}{Yongwei Wu}, \bibinfo{person}{Xuehai Qian},
  \bibinfo{person}{Kang Chen}, {and} \bibinfo{person}{Weimin Zheng}.}
  \bibinfo{year}{2018}\natexlab{}.
\newblock \showarticletitle{{CLIP}: A Disk {I/O} Focused Parallel Out-of-Core
  Graph Processing System}.
\newblock \bibinfo{journal}{\emph{IEEE Transactions on Parallel and Distributed
  Systems}} \bibinfo{volume}{30}, \bibinfo{number}{1} (\bibinfo{year}{2018}),
  \bibinfo{pages}{45--62}.
\newblock


\bibitem[\protect\citeauthoryear{Backstrom, Boldi, Rosa, Ugander, and
  Vigna}{Backstrom et~al\mbox{.}}{2012}]%
        {backstrom2012four}
\bibfield{author}{\bibinfo{person}{Lars Backstrom}, \bibinfo{person}{Paolo
  Boldi}, \bibinfo{person}{Marco Rosa}, \bibinfo{person}{Johan Ugander}, {and}
  \bibinfo{person}{Sebastiano Vigna}.} \bibinfo{year}{2012}\natexlab{}.
\newblock \showarticletitle{Four degrees of separation}. In
  \bibinfo{booktitle}{\emph{Proceedings of the 4th Annual ACM Web Science
  Conference}}. \bibinfo{pages}{33--42}.
\newblock


\bibitem[\protect\citeauthoryear{Boldi, Marino, Santini, and Vigna}{Boldi
  et~al\mbox{.}}{2014}]%
        {BMSB}
\bibfield{author}{\bibinfo{person}{Paolo Boldi}, \bibinfo{person}{Andrea
  Marino}, \bibinfo{person}{Massimo Santini}, {and} \bibinfo{person}{Sebastiano
  Vigna}.} \bibinfo{year}{2014}\natexlab{}.
\newblock \showarticletitle{{BUbiNG}: Massive Crawling for the Masses}. In
  \bibinfo{booktitle}{\emph{Proceedings of the Companion Publication of the
  23rd International Conference on World Wide Web}}.
  \bibinfo{publisher}{International World Wide Web Conferences Steering
  Committee}, \bibinfo{pages}{227--228}.
\newblock


\bibitem[\protect\citeauthoryear{Boldi, Rosa, Santini, and Vigna}{Boldi
  et~al\mbox{.}}{2011}]%
        {BRSLLP}
\bibfield{author}{\bibinfo{person}{Paolo Boldi}, \bibinfo{person}{Marco Rosa},
  \bibinfo{person}{Massimo Santini}, {and} \bibinfo{person}{Sebastiano Vigna}.}
  \bibinfo{year}{2011}\natexlab{}.
\newblock \showarticletitle{Layered Label Propagation: A MultiResolution
  Coordinate-Free Ordering for Compressing Social Networks}. In
  \bibinfo{booktitle}{\emph{Proceedings of the 20th international conference on
  World Wide Web}}, \bibfield{editor}{\bibinfo{person}{Sadagopan Srinivasan},
  \bibinfo{person}{Krithi Ramamritham}, \bibinfo{person}{Arun Kumar},
  \bibinfo{person}{M.~P. Ravindra}, \bibinfo{person}{Elisa Bertino}, {and}
  \bibinfo{person}{Ravi Kumar}} (Eds.). \bibinfo{publisher}{ACM Press},
  \bibinfo{pages}{587--596}.
\newblock


\bibitem[\protect\citeauthoryear{Boldi and Vigna}{Boldi and Vigna}{2004}]%
        {BoVWFI}
\bibfield{author}{\bibinfo{person}{Paolo Boldi} {and}
  \bibinfo{person}{Sebastiano Vigna}.} \bibinfo{year}{2004}\natexlab{}.
\newblock \showarticletitle{The {W}eb{G}raph Framework {I}: {C}ompression
  Techniques}. In \bibinfo{booktitle}{\emph{Proc. of the Thirteenth
  International World Wide Web Conference (WWW 2004)}}. \bibinfo{publisher}{ACM
  Press}, \bibinfo{address}{Manhattan, USA}, \bibinfo{pages}{595--601}.
\newblock


\bibitem[\protect\citeauthoryear{Boncz}{Boncz}{2013}]%
        {boncz2013ldbc}
\bibfield{author}{\bibinfo{person}{Peter Boncz}.}
  \bibinfo{year}{2013}\natexlab{}.
\newblock \showarticletitle{{LDBC}: benchmarks for graph and RDF data
  management}. In \bibinfo{booktitle}{\emph{Proceedings of the 17th
  International Database Engineering \& Applications Symposium}}.
  \bibinfo{pages}{1--2}.
\newblock


\bibitem[\protect\citeauthoryear{Borthakur}{Borthakur}{2007}]%
        {HDFS}
\bibfield{author}{\bibinfo{person}{Dhruba Borthakur}.}
  \bibinfo{year}{2007}\natexlab{}.
\newblock \showarticletitle{The hadoop distributed file system: Architecture
  and design}.
\newblock \bibinfo{journal}{\emph{Hadoop Project Website}}
  \bibinfo{volume}{11}, \bibinfo{number}{2007} (\bibinfo{year}{2007}),
  \bibinfo{pages}{21}.
\newblock


\bibitem[\protect\citeauthoryear{Bu, Borkar, Jia, Carey, and Condie}{Bu
  et~al\mbox{.}}{2014}]%
        {pregelix}
\bibfield{author}{\bibinfo{person}{Yingyi Bu}, \bibinfo{person}{Vinayak
  Borkar}, \bibinfo{person}{Jianfeng Jia}, \bibinfo{person}{Michael~J Carey},
  {and} \bibinfo{person}{Tyson Condie}.} \bibinfo{year}{2014}\natexlab{}.
\newblock \showarticletitle{Pregelix: Big (ger) graph analytics on a dataflow
  engine}.
\newblock \bibinfo{journal}{\emph{arXiv preprint arXiv:1407.0455}}
  (\bibinfo{year}{2014}).
\newblock


\bibitem[\protect\citeauthoryear{Chakrabarti, Zhan, and Faloutsos}{Chakrabarti
  et~al\mbox{.}}{2004}]%
        {chakrabarti2004r}
\bibfield{author}{\bibinfo{person}{Deepayan Chakrabarti},
  \bibinfo{person}{Yiping Zhan}, {and} \bibinfo{person}{Christos Faloutsos}.}
  \bibinfo{year}{2004}\natexlab{}.
\newblock \showarticletitle{{R-MAT}: A recursive model for graph mining}. In
  \bibinfo{booktitle}{\emph{Proceedings of the 2004 SIAM International
  Conference on Data Mining}}. SIAM, \bibinfo{pages}{442--446}.
\newblock


\bibitem[\protect\citeauthoryear{Chen, Ding, Wang, Chen, Zang, and Guan}{Chen
  et~al\mbox{.}}{2014}]%
        {Cyclops}
\bibfield{author}{\bibinfo{person}{Rong Chen}, \bibinfo{person}{Xin Ding},
  \bibinfo{person}{Peng Wang}, \bibinfo{person}{Haibo Chen},
  \bibinfo{person}{Binyu Zang}, {and} \bibinfo{person}{Haibing Guan}.}
  \bibinfo{year}{2014}\natexlab{}.
\newblock \showarticletitle{Computation and communication efficient graph
  processing with distributed immutable view}. In
  \bibinfo{booktitle}{\emph{Proceedings of the 23rd international symposium on
  High-performance parallel and distributed computing}}. ACM,
  \bibinfo{pages}{215--226}.
\newblock


\bibitem[\protect\citeauthoryear{Chen, Shi, Chen, Zang, Guan, and Chen}{Chen
  et~al\mbox{.}}{2019}]%
        {PowerLyra}
\bibfield{author}{\bibinfo{person}{Rong Chen}, \bibinfo{person}{Jiaxin Shi},
  \bibinfo{person}{Yanzhe Chen}, \bibinfo{person}{Binyu Zang},
  \bibinfo{person}{Haibing Guan}, {and} \bibinfo{person}{Haibo Chen}.}
  \bibinfo{year}{2019}\natexlab{}.
\newblock \showarticletitle{PowerLyra: Differentiated graph computation and
  partitioning on skewed graphs}.
\newblock \bibinfo{journal}{\emph{ACM Transactions on Parallel Computing
  (TOPC)}} \bibinfo{volume}{5}, \bibinfo{number}{3} (\bibinfo{year}{2019}),
  \bibinfo{pages}{13}.
\newblock


\bibitem[\protect\citeauthoryear{Chi, Dai, Wang, Sun, Li, and Yang}{Chi
  et~al\mbox{.}}{2016}]%
        {NXGraph}
\bibfield{author}{\bibinfo{person}{Yuze Chi}, \bibinfo{person}{Guohao Dai},
  \bibinfo{person}{Yu Wang}, \bibinfo{person}{Guangyu Sun},
  \bibinfo{person}{Guoliang Li}, {and} \bibinfo{person}{Huazhong Yang}.}
  \bibinfo{year}{2016}\natexlab{}.
\newblock \showarticletitle{NXgraph: An efficient graph processing system on a
  single machine}. In \bibinfo{booktitle}{\emph{2016 IEEE 32nd International
  Conference on Data Engineering (ICDE)}}. IEEE, \bibinfo{pages}{409--420}.
\newblock


\bibitem[\protect\citeauthoryear{Gonzalez, Low, Gu, Bickson, and
  Guestrin}{Gonzalez et~al\mbox{.}}{2012}]%
        {PowerGraph}
\bibfield{author}{\bibinfo{person}{Joseph~E Gonzalez}, \bibinfo{person}{Yucheng
  Low}, \bibinfo{person}{Haijie Gu}, \bibinfo{person}{Danny Bickson}, {and}
  \bibinfo{person}{Carlos Guestrin}.} \bibinfo{year}{2012}\natexlab{}.
\newblock \showarticletitle{PowerGraph: Distributed graph-parallel computation
  on natural graphs}. In \bibinfo{booktitle}{\emph{Presented as part of the
  10th {USENIX} Symposium on Operating Systems Design and Implementation
  ({OSDI} 12)}}. \bibinfo{pages}{17--30}.
\newblock


\bibitem[\protect\citeauthoryear{Gonzalez, Xin, Dave, Crankshaw, Franklin, and
  Stoica}{Gonzalez et~al\mbox{.}}{2014}]%
        {GraphX}
\bibfield{author}{\bibinfo{person}{Joseph~E Gonzalez},
  \bibinfo{person}{Reynold~S Xin}, \bibinfo{person}{Ankur Dave},
  \bibinfo{person}{Daniel Crankshaw}, \bibinfo{person}{Michael~J Franklin},
  {and} \bibinfo{person}{Ion Stoica}.} \bibinfo{year}{2014}\natexlab{}.
\newblock \showarticletitle{GraphX: Graph processing in a distributed dataflow
  framework}. In \bibinfo{booktitle}{\emph{11th {USENIX} Symposium on Operating
  Systems Design and Implementation ({OSDI} 14)}}. \bibinfo{pages}{599--613}.
\newblock


\bibitem[\protect\citeauthoryear{Grover and Leskovec}{Grover and
  Leskovec}{2016}]%
        {grover2016node2vec}
\bibfield{author}{\bibinfo{person}{Aditya Grover} {and} \bibinfo{person}{Jure
  Leskovec}.} \bibinfo{year}{2016}\natexlab{}.
\newblock \showarticletitle{node2vec: Scalable feature learning for networks}.
  In \bibinfo{booktitle}{\emph{Proceedings of the 22nd ACM SIGKDD international
  conference on Knowledge discovery and data mining}}.
  \bibinfo{pages}{855--864}.
\newblock


\bibitem[\protect\citeauthoryear{Han, Lee, Park, Lee, Kim, Kim, and Yu}{Han
  et~al\mbox{.}}{2013}]%
        {TurboGraph}
\bibfield{author}{\bibinfo{person}{Wook-Shin Han}, \bibinfo{person}{Sangyeon
  Lee}, \bibinfo{person}{Kyungyeol Park}, \bibinfo{person}{Jeong-Hoon Lee},
  \bibinfo{person}{Min-Soo Kim}, \bibinfo{person}{Jinha Kim}, {and}
  \bibinfo{person}{Hwanjo Yu}.} \bibinfo{year}{2013}\natexlab{}.
\newblock \showarticletitle{TurboGraph: a fast parallel graph engine handling
  billion-scale graphs in a single PC}. In
  \bibinfo{booktitle}{\emph{Proceedings of the 19th ACM SIGKDD international
  conference on Knowledge discovery and data mining}}. ACM,
  \bibinfo{pages}{77--85}.
\newblock


\bibitem[\protect\citeauthoryear{Ko and Han}{Ko and Han}{2018}]%
        {TurboGraph++}
\bibfield{author}{\bibinfo{person}{Seongyun Ko} {and}
  \bibinfo{person}{Wook-Shin Han}.} \bibinfo{year}{2018}\natexlab{}.
\newblock \showarticletitle{Turbograph++: A scalable and fast graph analytics
  system}. In \bibinfo{booktitle}{\emph{Proceedings of the 2018 International
  Conference on Management of Data}}. ACM, \bibinfo{pages}{395--410}.
\newblock


\bibitem[\protect\citeauthoryear{Kwak, Lee, Park, and Moon}{Kwak
  et~al\mbox{.}}{2010}]%
        {kwak2010twitter}
\bibfield{author}{\bibinfo{person}{Haewoon Kwak}, \bibinfo{person}{Changhyun
  Lee}, \bibinfo{person}{Hosung Park}, {and} \bibinfo{person}{Sue Moon}.}
  \bibinfo{year}{2010}\natexlab{}.
\newblock \showarticletitle{What is Twitter, a social network or a news
  media?}. In \bibinfo{booktitle}{\emph{Proceedings of the 19th international
  conference on World wide web}}. \bibinfo{pages}{591--600}.
\newblock


\bibitem[\protect\citeauthoryear{Kyrola, Blelloch, and Guestrin}{Kyrola
  et~al\mbox{.}}{2012}]%
        {GraphChi}
\bibfield{author}{\bibinfo{person}{Aapo Kyrola}, \bibinfo{person}{Guy
  Blelloch}, {and} \bibinfo{person}{Carlos Guestrin}.}
  \bibinfo{year}{2012}\natexlab{}.
\newblock \showarticletitle{GraphChi: Large-Scale Graph Computation on Just a
  {PC}}. In \bibinfo{booktitle}{\emph{Presented as part of the 10th {USENIX}
  Symposium on Operating Systems Design and Implementation ({OSDI} 12)}}.
  \bibinfo{pages}{31--46}.
\newblock


\bibitem[\protect\citeauthoryear{Lee, Kim, Lim, Noh, and Seo}{Lee
  et~al\mbox{.}}{2019}]%
        {lee2019pre}
\bibfield{author}{\bibinfo{person}{Eunjae Lee}, \bibinfo{person}{Junghyun Kim},
  \bibinfo{person}{Keunhak Lim}, \bibinfo{person}{Sam~H Noh}, {and}
  \bibinfo{person}{Jiwon Seo}.} \bibinfo{year}{2019}\natexlab{}.
\newblock \showarticletitle{Pre-select static caching and neighborhood ordering
  for BFS-like algorithms on disk-based graph engines}. In
  \bibinfo{booktitle}{\emph{2019 {USENIX} Annual Technical Conference ({USENIX}
  {ATC} 19)}}. \bibinfo{pages}{459--474}.
\newblock


\bibitem[\protect\citeauthoryear{Leskovec, Chakrabarti, Kleinberg, and
  Faloutsos}{Leskovec et~al\mbox{.}}{2005}]%
        {leskovec2005realistic}
\bibfield{author}{\bibinfo{person}{Jurij Leskovec}, \bibinfo{person}{Deepayan
  Chakrabarti}, \bibinfo{person}{Jon Kleinberg}, {and}
  \bibinfo{person}{Christos Faloutsos}.} \bibinfo{year}{2005}\natexlab{}.
\newblock \showarticletitle{Realistic, mathematically tractable graph
  generation and evolution, using kronecker multiplication}. In
  \bibinfo{booktitle}{\emph{European conference on principles of data mining
  and knowledge discovery}}. Springer, \bibinfo{pages}{133--145}.
\newblock


\bibitem[\protect\citeauthoryear{Lin, Zhu, Yu, Tang, Xue, Chen, Zhang, Hoefler,
  Ma, Liu, et~al\mbox{.}}{Lin et~al\mbox{.}}{2018}]%
        {ShenTu}
\bibfield{author}{\bibinfo{person}{Heng Lin}, \bibinfo{person}{Xiaowei Zhu},
  \bibinfo{person}{Bowen Yu}, \bibinfo{person}{Xiongchao Tang},
  \bibinfo{person}{Wei Xue}, \bibinfo{person}{Wenguang Chen},
  \bibinfo{person}{Lufei Zhang}, \bibinfo{person}{Torsten Hoefler},
  \bibinfo{person}{Xiaosong Ma}, \bibinfo{person}{Xin Liu}, {et~al\mbox{.}}}
  \bibinfo{year}{2018}\natexlab{}.
\newblock \showarticletitle{ShenTu: processing multi-trillion edge graphs on
  millions of cores in seconds}. In \bibinfo{booktitle}{\emph{Proceedings of
  the International Conference for High Performance Computing, Networking,
  Storage, and Analysis}}. IEEE Press, \bibinfo{pages}{56}.
\newblock


\bibitem[\protect\citeauthoryear{Liu and Huang}{Liu and Huang}{2017}]%
        {Graphene}
\bibfield{author}{\bibinfo{person}{Hang Liu} {and} \bibinfo{person}{H~Howie
  Huang}.} \bibinfo{year}{2017}\natexlab{}.
\newblock \showarticletitle{Graphene: Fine-grained {IO} management for graph
  computing}. In \bibinfo{booktitle}{\emph{15th {USENIX} Conference on File and
  Storage Technologies ({FAST} 17)}}. \bibinfo{pages}{285--300}.
\newblock


\bibitem[\protect\citeauthoryear{Low, Bickson, Gonzalez, Guestrin, Kyrola, and
  Hellerstein}{Low et~al\mbox{.}}{2012}]%
        {GraphLab}
\bibfield{author}{\bibinfo{person}{Yucheng Low}, \bibinfo{person}{Danny
  Bickson}, \bibinfo{person}{Joseph Gonzalez}, \bibinfo{person}{Carlos
  Guestrin}, \bibinfo{person}{Aapo Kyrola}, {and} \bibinfo{person}{Joseph~M
  Hellerstein}.} \bibinfo{year}{2012}\natexlab{}.
\newblock \showarticletitle{Distributed GraphLab: a framework for machine
  learning and data mining in the cloud}.
\newblock \bibinfo{journal}{\emph{Proceedings of the VLDB Endowment}}
  \bibinfo{volume}{5}, \bibinfo{number}{8} (\bibinfo{year}{2012}),
  \bibinfo{pages}{716--727}.
\newblock


\bibitem[\protect\citeauthoryear{Maass, Min, Kashyap, Kang, Kumar, and
  Kim}{Maass et~al\mbox{.}}{2017}]%
        {Mosaic}
\bibfield{author}{\bibinfo{person}{Steffen Maass}, \bibinfo{person}{Changwoo
  Min}, \bibinfo{person}{Sanidhya Kashyap}, \bibinfo{person}{Woonhak Kang},
  \bibinfo{person}{Mohan Kumar}, {and} \bibinfo{person}{Taesoo Kim}.}
  \bibinfo{year}{2017}\natexlab{}.
\newblock \showarticletitle{Mosaic: Processing a trillion-edge graph on a
  single machine}. In \bibinfo{booktitle}{\emph{Proceedings of the Twelfth
  European Conference on Computer Systems}}. ACM, \bibinfo{pages}{527--543}.
\newblock


\bibitem[\protect\citeauthoryear{Malewicz, Austern, Bik, Dehnert, Horn, Leiser,
  and Czajkowski}{Malewicz et~al\mbox{.}}{2010}]%
        {Pregel}
\bibfield{author}{\bibinfo{person}{Grzegorz Malewicz},
  \bibinfo{person}{Matthew~H Austern}, \bibinfo{person}{Aart~JC Bik},
  \bibinfo{person}{James~C Dehnert}, \bibinfo{person}{Ilan Horn},
  \bibinfo{person}{Naty Leiser}, {and} \bibinfo{person}{Grzegorz Czajkowski}.}
  \bibinfo{year}{2010}\natexlab{}.
\newblock \showarticletitle{Pregel: a system for large-scale graph processing}.
  In \bibinfo{booktitle}{\emph{Proceedings of the 2010 ACM SIGMOD International
  Conference on Management of data}}. ACM, \bibinfo{pages}{135--146}.
\newblock


\bibitem[\protect\citeauthoryear{McCune, Weninger, and Madey}{McCune
  et~al\mbox{.}}{2015}]%
        {Thinking}
\bibfield{author}{\bibinfo{person}{Robert~Ryan McCune}, \bibinfo{person}{Tim
  Weninger}, {and} \bibinfo{person}{Greg Madey}.}
  \bibinfo{year}{2015}\natexlab{}.
\newblock \showarticletitle{Thinking like a vertex: a survey of vertex-centric
  frameworks for large-scale distributed graph processing}.
\newblock \bibinfo{journal}{\emph{ACM Computing Surveys (CSUR)}}
  \bibinfo{volume}{48}, \bibinfo{number}{2} (\bibinfo{year}{2015}),
  \bibinfo{pages}{25}.
\newblock


\bibitem[\protect\citeauthoryear{McSherry, Isard, and Murray}{McSherry
  et~al\mbox{.}}{2015}]%
        {mcsherry2015scalability}
\bibfield{author}{\bibinfo{person}{Frank McSherry}, \bibinfo{person}{Michael
  Isard}, {and} \bibinfo{person}{Derek~G Murray}.}
  \bibinfo{year}{2015}\natexlab{}.
\newblock \showarticletitle{Scalability! {B}ut at what {COST}?}. In
  \bibinfo{booktitle}{\emph{15th Workshop on Hot Topics in Operating Systems
  (HotOS {XV})}}.
\newblock


\bibitem[\protect\citeauthoryear{Nguyen, Lenharth, and Pingali}{Nguyen
  et~al\mbox{.}}{2013}]%
        {Galois}
\bibfield{author}{\bibinfo{person}{Donald Nguyen}, \bibinfo{person}{Andrew
  Lenharth}, {and} \bibinfo{person}{Keshav Pingali}.}
  \bibinfo{year}{2013}\natexlab{}.
\newblock \showarticletitle{A lightweight infrastructure for graph analytics}.
  In \bibinfo{booktitle}{\emph{Proceedings of the Twenty-Fourth ACM Symposium
  on Operating Systems Principles}}. ACM, \bibinfo{pages}{456--471}.
\newblock


\bibitem[\protect\citeauthoryear{Page, Brin, Motwani, and Winograd}{Page
  et~al\mbox{.}}{1999}]%
        {page1999pagerank}
\bibfield{author}{\bibinfo{person}{Lawrence Page}, \bibinfo{person}{Sergey
  Brin}, \bibinfo{person}{Rajeev Motwani}, {and} \bibinfo{person}{Terry
  Winograd}.} \bibinfo{year}{1999}\natexlab{}.
\newblock \bibinfo{booktitle}{\emph{The pagerank citation ranking: Bringing
  order to the web.}}
\newblock \bibinfo{type}{{T}echnical {R}eport}. \bibinfo{institution}{Stanford
  InfoLab}.
\newblock


\bibitem[\protect\citeauthoryear{Roy, Bindschaedler, Malicevic, and
  Zwaenepoel}{Roy et~al\mbox{.}}{2015}]%
        {Chaos}
\bibfield{author}{\bibinfo{person}{Amitabha Roy}, \bibinfo{person}{Laurent
  Bindschaedler}, \bibinfo{person}{Jasmina Malicevic}, {and}
  \bibinfo{person}{Willy Zwaenepoel}.} \bibinfo{year}{2015}\natexlab{}.
\newblock \showarticletitle{Chaos: Scale-out graph processing from secondary
  storage}. In \bibinfo{booktitle}{\emph{Proceedings of the 25th Symposium on
  Operating Systems Principles}}. ACM, \bibinfo{pages}{410--424}.
\newblock


\bibitem[\protect\citeauthoryear{Roy, Mihailovic, and Zwaenepoel}{Roy
  et~al\mbox{.}}{2013}]%
        {XStream}
\bibfield{author}{\bibinfo{person}{Amitabha Roy}, \bibinfo{person}{Ivo
  Mihailovic}, {and} \bibinfo{person}{Willy Zwaenepoel}.}
  \bibinfo{year}{2013}\natexlab{}.
\newblock \showarticletitle{X-Stream: Edge-centric graph processing using
  streaming partitions}. In \bibinfo{booktitle}{\emph{Proceedings of the
  Twenty-Fourth ACM Symposium on Operating Systems Principles}}. ACM,
  \bibinfo{pages}{472--488}.
\newblock


\bibitem[\protect\citeauthoryear{Shun and Blelloch}{Shun and Blelloch}{2013}]%
        {Ligra}
\bibfield{author}{\bibinfo{person}{Julian Shun} {and} \bibinfo{person}{Guy~E
  Blelloch}.} \bibinfo{year}{2013}\natexlab{}.
\newblock \showarticletitle{Ligra: a lightweight graph processing framework for
  shared memory}. In \bibinfo{booktitle}{\emph{ACM Sigplan Notices}},
  Vol.~\bibinfo{volume}{48}. ACM, \bibinfo{pages}{135--146}.
\newblock


\bibitem[\protect\citeauthoryear{Shun, Dhulipala, and Blelloch}{Shun
  et~al\mbox{.}}{2015}]%
        {Ligra+}
\bibfield{author}{\bibinfo{person}{Julian Shun}, \bibinfo{person}{Laxman
  Dhulipala}, {and} \bibinfo{person}{Guy~E Blelloch}.}
  \bibinfo{year}{2015}\natexlab{}.
\newblock \showarticletitle{Smaller and faster: Parallel processing of
  compressed graphs with {L}igra+}. In \bibinfo{booktitle}{\emph{2015 Data
  Compression Conference}}. IEEE, \bibinfo{pages}{403--412}.
\newblock


\bibitem[\protect\citeauthoryear{Vora}{Vora}{2019}]%
        {vora2019lumos}
\bibfield{author}{\bibinfo{person}{Keval Vora}.}
  \bibinfo{year}{2019}\natexlab{}.
\newblock \showarticletitle{{LUMOS}: Dependency-Driven Disk-based Graph
  Processing}. In \bibinfo{booktitle}{\emph{2019 {USENIX} Annual Technical
  Conference ({USENIX} {ATC} 19)}}. \bibinfo{pages}{429--442}.
\newblock


\bibitem[\protect\citeauthoryear{Vora, Xu, and Gupta}{Vora
  et~al\mbox{.}}{2016}]%
        {vora2016load}
\bibfield{author}{\bibinfo{person}{Keval Vora}, \bibinfo{person}{Guoqing Xu},
  {and} \bibinfo{person}{Rajiv Gupta}.} \bibinfo{year}{2016}\natexlab{}.
\newblock \showarticletitle{Load the edges you need: A generic {I/O}
  optimization for disk-based graph processing}. In
  \bibinfo{booktitle}{\emph{2016 {USENIX} Annual Technical Conference ({USENIX}
  {ATC} 16)}}. \bibinfo{pages}{507--522}.
\newblock


\bibitem[\protect\citeauthoryear{Wang, Gu, Bao, Yu, and Yu}{Wang
  et~al\mbox{.}}{2016}]%
        {HybridGraph}
\bibfield{author}{\bibinfo{person}{Zhigang Wang}, \bibinfo{person}{Yu Gu},
  \bibinfo{person}{Yubin Bao}, \bibinfo{person}{Ge Yu}, {and}
  \bibinfo{person}{Jeffrey~Xu Yu}.} \bibinfo{year}{2016}\natexlab{}.
\newblock \showarticletitle{Hybrid pulling/pushing for {I/O}-efficient
  distributed and iterative graph computing}. In
  \bibinfo{booktitle}{\emph{Proceedings of the 2016 International Conference on
  Management of Data}}. ACM, \bibinfo{pages}{479--494}.
\newblock


\bibitem[\protect\citeauthoryear{Wu, Yang, Xue, Xiao, Miao, Wei, Lin, Dai, and
  Zhou}{Wu et~al\mbox{.}}{2015}]%
        {GraM}
\bibfield{author}{\bibinfo{person}{Ming Wu}, \bibinfo{person}{Fan Yang},
  \bibinfo{person}{Jilong Xue}, \bibinfo{person}{Wencong Xiao},
  \bibinfo{person}{Youshan Miao}, \bibinfo{person}{Lan Wei},
  \bibinfo{person}{Haoxiang Lin}, \bibinfo{person}{Yafei Dai}, {and}
  \bibinfo{person}{Lidong Zhou}.} \bibinfo{year}{2015}\natexlab{}.
\newblock \showarticletitle{{GraM}: scaling graph computation to the
  trillions}. In \bibinfo{booktitle}{\emph{Proceedings of the Sixth ACM
  Symposium on Cloud Computing}}. ACM, \bibinfo{pages}{408--421}.
\newblock


\bibitem[\protect\citeauthoryear{Xiao, Xue, Miao, Li, Chen, Wu, Li, and
  Zhou}{Xiao et~al\mbox{.}}{2017}]%
        {xiao2017tux2}
\bibfield{author}{\bibinfo{person}{Wencong Xiao}, \bibinfo{person}{Jilong Xue},
  \bibinfo{person}{Youshan Miao}, \bibinfo{person}{Zhen Li},
  \bibinfo{person}{Cheng Chen}, \bibinfo{person}{Ming Wu}, \bibinfo{person}{Wei
  Li}, {and} \bibinfo{person}{Lidong Zhou}.} \bibinfo{year}{2017}\natexlab{}.
\newblock \showarticletitle{Tux$^2$: Distributed Graph Computation for Machine
  Learning}. In \bibinfo{booktitle}{\emph{14th {USENIX} Symposium on Networked
  Systems Design and Implementation ({NSDI} 17)}}. \bibinfo{pages}{669--682}.
\newblock


\bibitem[\protect\citeauthoryear{Xie, Chen, Guan, Zang, and Chen}{Xie
  et~al\mbox{.}}{2015}]%
        {PowerSwitch}
\bibfield{author}{\bibinfo{person}{Chenning Xie}, \bibinfo{person}{Rong Chen},
  \bibinfo{person}{Haibing Guan}, \bibinfo{person}{Binyu Zang}, {and}
  \bibinfo{person}{Haibo Chen}.} \bibinfo{year}{2015}\natexlab{}.
\newblock \showarticletitle{Sync or async: Time to fuse for distributed
  graph-parallel computation}. In \bibinfo{booktitle}{\emph{ACM SIGPLAN
  Notices}}, Vol.~\bibinfo{volume}{50}. ACM, \bibinfo{pages}{194--204}.
\newblock


\bibitem[\protect\citeauthoryear{Yan, Huang, Liu, Chen, Cheng, Wu, and
  Zhang}{Yan et~al\mbox{.}}{2017}]%
        {graphd}
\bibfield{author}{\bibinfo{person}{Da Yan}, \bibinfo{person}{Yuzhen Huang},
  \bibinfo{person}{Miao Liu}, \bibinfo{person}{Hongzhi Chen},
  \bibinfo{person}{James Cheng}, \bibinfo{person}{Huanhuan Wu}, {and}
  \bibinfo{person}{Chengcui Zhang}.} \bibinfo{year}{2017}\natexlab{}.
\newblock \showarticletitle{Graphd: Distributed vertex-centric graph processing
  beyond the memory limit}.
\newblock \bibinfo{journal}{\emph{IEEE Transactions on Parallel and Distributed
  Systems}} \bibinfo{volume}{29}, \bibinfo{number}{1} (\bibinfo{year}{2017}),
  \bibinfo{pages}{99--114}.
\newblock


\bibitem[\protect\citeauthoryear{Zaharia, Chowdhury, Franklin, Shenker, and
  Stoica}{Zaharia et~al\mbox{.}}{2010}]%
        {Spark}
\bibfield{author}{\bibinfo{person}{Matei Zaharia}, \bibinfo{person}{Mosharaf
  Chowdhury}, \bibinfo{person}{Michael~J Franklin}, \bibinfo{person}{Scott
  Shenker}, {and} \bibinfo{person}{Ion Stoica}.}
  \bibinfo{year}{2010}\natexlab{}.
\newblock \showarticletitle{Spark: Cluster computing with working sets.}
\newblock \bibinfo{journal}{\emph{HotCloud}} \bibinfo{volume}{10},
  \bibinfo{number}{10-10} (\bibinfo{year}{2010}), \bibinfo{pages}{95}.
\newblock


\bibitem[\protect\citeauthoryear{Zhang, Chen, and Chen}{Zhang
  et~al\mbox{.}}{2015}]%
        {Polymer}
\bibfield{author}{\bibinfo{person}{Kaiyuan Zhang}, \bibinfo{person}{Rong Chen},
  {and} \bibinfo{person}{Haibo Chen}.} \bibinfo{year}{2015}\natexlab{}.
\newblock \showarticletitle{NUMA-aware graph-structured analytics}.
\newblock \bibinfo{journal}{\emph{ACM SIGPLAN Notices}} \bibinfo{volume}{50},
  \bibinfo{number}{8} (\bibinfo{year}{2015}), \bibinfo{pages}{183--193}.
\newblock


\bibitem[\protect\citeauthoryear{Zhao, Ding, Liu, Yu, Han, and Feng}{Zhao
  et~al\mbox{.}}{2019}]%
        {zhao2019cacheap}
\bibfield{author}{\bibinfo{person}{Peng Zhao}, \bibinfo{person}{Chen Ding},
  \bibinfo{person}{Lei Liu}, \bibinfo{person}{Jiping Yu},
  \bibinfo{person}{Wentao Han}, {and} \bibinfo{person}{Xiao-Bing Feng}.}
  \bibinfo{year}{2019}\natexlab{}.
\newblock \showarticletitle{Cacheap: Portable and Collaborative {I/O}
  Optimization for Graph Processing}.
\newblock \bibinfo{journal}{\emph{Journal of Computer Science and Technology}}
  \bibinfo{volume}{34}, \bibinfo{number}{3} (\bibinfo{year}{2019}),
  \bibinfo{pages}{690--706}.
\newblock


\bibitem[\protect\citeauthoryear{Zheng, Mhembere, Burns, Vogelstein, Priebe,
  and Szalay}{Zheng et~al\mbox{.}}{2015}]%
        {FlashGraph}
\bibfield{author}{\bibinfo{person}{Da Zheng}, \bibinfo{person}{Disa Mhembere},
  \bibinfo{person}{Randal Burns}, \bibinfo{person}{Joshua Vogelstein},
  \bibinfo{person}{Carey~E Priebe}, {and} \bibinfo{person}{Alexander~S
  Szalay}.} \bibinfo{year}{2015}\natexlab{}.
\newblock \showarticletitle{{FlashGraph}: Processing billion-node graphs on an
  array of commodity SSDs}. In \bibinfo{booktitle}{\emph{13th {USENIX}
  Conference on File and Storage Technologies ({FAST} 15)}}.
  \bibinfo{pages}{45--58}.
\newblock


\bibitem[\protect\citeauthoryear{Zhu, Chen, Zheng, and Ma}{Zhu
  et~al\mbox{.}}{2016}]%
        {Gemini}
\bibfield{author}{\bibinfo{person}{Xiaowei Zhu}, \bibinfo{person}{Wenguang
  Chen}, \bibinfo{person}{Weimin Zheng}, {and} \bibinfo{person}{Xiaosong Ma}.}
  \bibinfo{year}{2016}\natexlab{}.
\newblock \showarticletitle{Gemini: A computation-centric distributed graph
  processing system}. In \bibinfo{booktitle}{\emph{12th {USENIX} Symposium on
  Operating Systems Design and Implementation ({OSDI} 16)}}.
  \bibinfo{pages}{301--316}.
\newblock


\bibitem[\protect\citeauthoryear{Zhu, Han, and Chen}{Zhu et~al\mbox{.}}{2015}]%
        {GridGraph}
\bibfield{author}{\bibinfo{person}{Xiaowei Zhu}, \bibinfo{person}{Wentao Han},
  {and} \bibinfo{person}{Wenguang Chen}.} \bibinfo{year}{2015}\natexlab{}.
\newblock \showarticletitle{{GridGraph}: Large-scale graph processing on a
  single machine using 2-level hierarchical partitioning}. In
  \bibinfo{booktitle}{\emph{2015 {USENIX} Annual Technical Conference ({USENIX}
  {ATC} 15)}}. \bibinfo{pages}{375--386}.
\newblock


\end{thebibliography}



\end{document}